\documentclass[letterpaper,12pt]{article}
\usepackage[margin=1in]{geometry}
\usepackage{amsmath,amsfonts,amssymb}
\usepackage[normalem]{ulem}
\usepackage{enumerate}
\usepackage[title,toc,titletoc]{appendix}
\usepackage{graphicx}
\usepackage{setspace}
\usepackage{subcaption}
\usepackage{multirow}
\usepackage{floatrow, float}
\usepackage{hyperref}
\usepackage{bm}
\usepackage{bbm}
\usepackage{wrapfig}
\usepackage{cuted}
\usepackage[linesnumbered,ruled,vlined]{algorithm2e}

\newfloatcommand{capbtabbox}{table}[][\FBwidth]

\usepackage{natbib}

\usepackage{graphicx}
\usepackage{float}
\graphicspath{{./figures/}}

\def\spacingset#1{\renewcommand{\baselinestretch}%
	{#1}\small\normalsize} \spacingset{1}

\usepackage[dvipsnames]{xcolor}

\usepackage{url}

\usepackage{bm} 
\newcommand{\phib}{\bm{\phi}}

\newcommand{\thetab}{\bm{\theta}}  

\newcommand{\db}{\bm{d}}

\newcommand{\gb}{\bm{g}}
\newcommand{\Gb}{\bm{G}}

\newcommand{\tb}{\bm{t}}

\newcommand{\yb}{\bm{y}}    

\newcommand{\zb}{\bm{z}}
\newcommand{\Omegab}{\bm{\Omega}} 

\newcommand{\lambdab}{\bm{\lambda}}

\newcommand{\gammab}{\bm{\gamma}}

\newcommand{\cN}{\mathcal{N}}   

\newcommand*\diff{\mathop{}\!\textrm{d}}

\DeclareMathOperator*{\argmin}{arg\,min}

\newtheorem{theorem}{Theorem}
\newtheorem{definition}{Definition}

\newtheorem{proposition}{Proposition}

\providecommand{\keywords}[1]
{
  \small	
  \textbf{\textit{Keywords---}} #1
}

\title{Variational Inference with Vine Copulas: An efficient Approach for Bayesian Computer Model Calibration}
\author{Vojtech Kejzlar$^{1}$ and Tapabrata Maiti$^{2}$ \\
\small $^{1}$Department of Mathematics and Statistics, Skidmore College \\
\small $^{2}$Department of Statistics and Probability, Michigan State University }

\date{}
\begin{document}
\maketitle
\begin{abstract}
With the advancements of computer architectures, the use of computational models proliferates to solve complex problems in many scientific applications such as nuclear physics and climate research. However, the potential of such models is often hindered because they tend to be computationally expensive and consequently ill-fitting for uncertainty quantification. Furthermore, they are usually not calibrated with real-time observations. We develop a computationally efficient algorithm based on variational Bayes inference (VBI) for calibration of computer models with Gaussian processes. Unfortunately, the speed and scalability of VBI diminishes when applied to the calibration framework with dependent data. To preserve the efficiency of VBI, we adopt a pairwise decomposition of the data likelihood using vine copulas that separate the information on dependence structure in data from their marginal distributions. We provide both theoretical and empirical evidence for the computational scalability of our methodology and describe all the necessary details for an efficient implementation of the proposed algorithm. We also demonstrate the opportunities given by our method for practitioners on a real data example through calibration of the Liquid Drop Model of nuclear binding energies.
\end{abstract}

\keywords{Bayesian inference; Computer experiments; Gaussian process; Nuclear physics; Vine copula; Uncertainty quantification; Prediction}

\section{Introduction}\label{sec:introduction}
The ever-growing access to high performance computing in scientific communities has enabled development of complex computer models in fields such as nuclear physics, climatology, and engineering that produce massive amounts of data. These models need real-time calibration with quantified uncertainties. Bayesian methodology combined with Gaussian process modeling has been heavily utilized for calibration of computer models due to its natural way to account for various sources of uncertainty; see \cite{HigdonJPG15}, and \cite{King19} for examples in nuclear physics, \cite{Sexton2012} and \cite{Pollard2016} for examples in climatology, and \cite{ williams2006}, \cite{Plumlee16} and \cite{Zhang2019} for applications in engineering. 

The framework for Bayesian calibration of computer models was developed by \cite{KoH} with extensions provided by \cite{Higdon04, Higdon08, Vlid, Plumee, Plumlee2019}, and \cite{MenGu18}, to name a few. Despite its popularity, however, Bayesian calibration becomes infeasible in big-data scenarios with complex and many-parameter models because it relies on Markov chain Monte Carlo (MCMC) algorithms to approximate posterior densities. 

This text presents a scalable and statistically principled approach to Bayesian calibration of computer models. We offer an alternative approximation to posterior densities using variational Bayesian inference (VBI), which originated as a machine learning algorithm that approximates a target density through optimization. Statisticians and computer scientists (starting with \cite{Peterson, Jordan1999}) have been widely using variational techniques because they tend to be faster and easier to scale to massive datasets. Moreover, the recently published frequentist consistency of variational Bayes by \cite{VIth} established VBI as a theoretically valid procedure. 
The scalability of VBI in modern applications hinges on efficiency of stochastic
optimization in scenarios with independent data points. This efficiency, however, diminishes in the
case of Bayesian calibration of computer models due to dependence structure in data \citep{robbins1951, Hoffman13a}.
To maintain the speed and scalability of VBI, we adopt a pairwise decomposition of data likelihood using vine copulas that separate the information on dependence structure in data from their marginal distributions \citep{Kurowicka2006}. Our specific contributions are as follows:
\begin{enumerate}
	\item We propose a novel version of the black-box variational inference \citep{Ranganath14} for Bayesian calibration of computer models that preserves the efficiency of stochastic optimization in scenario with dependent data. Python code with our algorithm is available at \url{https://github.com/kejzlarv/VBI_Calibration}.
	\item We incorporate Rao-Blackwellization, control variates, and importance sampling to reduce the variance of noisy gradient estimates involved in our algorithm.
	\item We provide both theoretical and empirical evidence for scalability of our methodology and establish its superiority over the Metropolis-Hastings algorithm and the No-U-Turn sampler both in terms of time efficiency and memory requirements.
	\item Finally, we demonstrate the opportunities in uncertainty quantification given by the proposed algorithm on a real-word example in the field of nuclear physics.
\end{enumerate} 

\subsection{Outline of this paper}
In Section \ref{sec:framework}, we describe the framework for Bayesian calibration of computer models and give an overview of VBI. In Section \ref{sec:vc}, we derive our proposed VBI approach to perform inexpensive and scalable calibration. We establish statistical validity of the method and provide theoretical justification for its scalability. Subsequently, in Section \ref{sec:implemenation}, we discuss the implementation details with focus on strategies to reduce the variance of the gradient. Section \ref{sec:applications} presents a simulation study comparing our approach with the state-of-the-art methods to approximate posterior distribution and illustrates our method on a real-data application.

\section{Background and Theoretical Framework}\label{sec:framework}
Formally, let $\yb = (y_1, \dots, y_n)$ be observations of a physical process $\zeta(\tb_i)$ depending on a known set of inputs $\tb_i \in \Omegab \subset \mathbb{R}^p$. Assume that $y_i$ follows
\begin{equation}\label{eqn:VBI:physicalProcess}
    y_i = \zeta(\tb_i) + \sigma \epsilon_i,
\end{equation}
where $\sigma$ represent the scale of observation error $\epsilon_i \stackrel{i.i.d.}{\sim} \cN(0,1)$. As a mathematical description of $\zeta$, we consider a computer model $f_m$ defined as the mapping $(\tb, \thetab)\mapsto f_m(\tb, \thetab)$ which depends on an additional set of inputs $\thetab \in \Theta \subset \mathbb{R}^{p'}$ that we call calibration parameters. These are fixed but unknown quantities representing fundamental properties of the physical process that cannot be directly measured or controlled in an experiment. Model calibration corresponds to determining the unknown and hypothetical \emph{true} value of the parameter $\thetab$, at which the physical process $\zeta(\tb)$ would satisfy $\zeta(\tb) = f(\tb,\thetab) + \delta(\tb)$; $\delta(\tb)$ is the systematic discrepancy of the model whose form is generally unknown. We assume a single value of calibration parameter $\thetab$ to be common among all the observations $y_i$ and all the future instances of the physical process.

Overall, we can write the complete statistical model as
\begin{equation}\label{eqn:completeModel}
y_i = f(\tb_i,\thetab) + \delta(\tb_i) + \epsilon_i.
\end{equation}
It is often the case that the evaluation of computer model $f_m$ is too expensive in terms of both time and space (memory). Common practice is to reduce the number of necessary computer model evaluations by considering a Gaussian process (GP) prior model:
\begin{equation*}
			f_m(\tb, \thetab) \sim \mathcal{GP}(m_f(\tb, \thetab), k_f((\tb,\thetab),(\tb', \thetab'))).
\end{equation*}
In this setup, the data also include set of model evaluations $\zb = (z_1,\dots, z_{s})$ over a grid $\{(\widetilde{\tb}_1, \widetilde{\thetab}_1), \dots, (\widetilde{\tb}_{s}, \widetilde{\thetab}_{s})\}$. These are usually selected sequentially using some space-filling design such us uniform or Latin hypercube design \citep{MORRIS1995381}, which is a design that has a good coverage of the space with evenly distributed points in each one-dimensional projection. The discrepancy function $\delta(\tb)$, while intrinsically deterministic, is also modeled by a GP. The complete dataset $\db$ consists of $n$ observations $y_i$ from the physical process $\zeta$ and $s$ evaluations $z_j$ of the computer model $f_m$, i.e. $\bm{d} = (d_1, \dots, d_{n+s}) := (\yb,\zb)$, and follows the multivariate normal distribution
\begin{equation}\label{eqn:VBI:dModel}
    \db | \phib \sim \cN(M(\phib), K(\phib)),
\end{equation}
where $\phib = (\thetab, \gammab, \sigma)$ is the set of all unknown parameters with $\gammab$ denoting the set of hyperparameters of the GPs' mean and covariance functions. $M(\phib)$ is the mean vector and $K(\phib)$ is the covariance matrix given by the GPs' specifications.

Under this framework, the Bayesian predictions of new values $\yb^*$ of a physical process $\zeta$ are given by the posterior predictive distribution $p(\yb^*|\db)$, namely
\begin{equation}\label{eqn:VBI:BayesPredictiveDensity} 
p(\yb^*|\db) = \int p(\yb^*| \db, \phib) p(\phib | \db) \diff \phib. 
\end{equation}
The conditional density $p(\yb^*| \db, \phib)$ is a multivariate normal density given by the statistical model \eqref{eqn:completeModel} and the specification of GPs. The posterior distribution of the unknown parameters $p(\phib | \db)$ is given by the Bayes' theorem. The term ``calibration" in the Bayesian paradigm includes both an estimation of $\phib$ and a full evaluation of uncertainty for every parameter under a prior uncertainty expressed by $p(\phib)$. It is also worth noting that the posterior predictive density is rarely computed directly from \eqref{eqn:VBI:BayesPredictiveDensity}. Instead, we first generate samples $\phib^{(1)}, \dots,\phib^{(M)}$ from $p(\phib| \db)$ and then obtain samples $\yb^{*(1)}, \dots, \yb^{*(M)}$ so that $\yb^{*(i)} \sim p(\yb^*| \db, \phib^{(i)})$, $i = 1, \dots, M$. The posterior predictive density is approximated using the empirical density of $\yb^{*(1)}, \dots, \yb^{*(M)}$.

As a consequence of this simple two-step algorithm, we are interested in effective sampling (approximation) from the posterior distribution $p(\phib|\bm{d})$. {\em This becomes quickly infeasible with increasing size of datasets, number of parameters, and model complexity}. Traditional MCMC methods that approximate $p(\phib|\bm{d})$---such as the Metropolis-Hastings (MH) algorithm \citep{Metropolis} or more advanced ones including Hamiltonian Monte Carlo or the No-U-Turn Sampler (NUTS) \citep{NUTS}---typically fail because of the computational costs associated with the evaluation of $p(\bm{d}|\phib)$. The standard approaches to scalable Bayesian inference are in general not applicable here because of the highly correlated structure of $K(\phib)$ or the nature of calibration itself. Indeed, parallelization of MCMC \citep{Neiswanger2014} works in the case of and independent datest $\bm{d}$, and GP approximation methods are developed in the context of regression problems \citep{QuinoneroCandela, pmlrv5titsias09a, Bauer2016}. We emphasize that our context is much more complex and that our approach is not focused on developing parallel computing, but rather exploiting probabilistic theory of approximation to reduce the computational cost.

\subsection{Variational Bayes Inference (VBI)}\label{sec:vbi}
VBI is an optimization based method that approximates $p(\phib|\bm{d})$ by a family of distributions $q(\phib|\lambdab)$ over latent variables with its own variational parameter $\lambdab$. Many commonly used families exist with the simplest mean-field family assuming independence of all the components in $\phib$; see \cite{MAL001,pmlr-v38-hoffman15, Ranganath:2016, Tran2015, Tran:2017} for examples of more sophisticated families. The approximate distribution $q^*$ is chosen to satisfy
\begin{equation}\label{eqn:VBI:KL}
q^* = \argmin_{q(\phib|\lambdab)} KL(q(\phib|\lambdab)||p(\phib|\bm{d})).
\end{equation}
Here, $KL$ denotes the Kullback-Leibler divergence of $q(\phib|\lambdab)$ from $p(\phib|\bm{d})$. Finding $q^*$ is done in practice by maximizing the \textit{evidence lower bound (ELBO)}
\begin{equation}\label{eqn:VBI:ELBO}
\mathcal{L}(\lambdab) = \mathbb{E}_q\bigg[\log p(\bm{d}|\phib)\bigg] - KL(q(\phib|\lambdab)||p(\phib)),
\end{equation}
which is a sum of the expected data log-likelihood and the negative $KL$ divergence between the combined prior distribution $p(\phib)$ of calibration parameters, the error scale $\sigma$, and GP hyperparameters and the variational distribution $q(\phib|\lambdab)$. Note that we set $\mathcal{L}(\lambdab) := \mathcal{L}(q(\phib|\lambdab))$ for the ease of notation. Minimizing the ELBO is equivalent to minimizing the original objective function.

The ELBO can be optimized via the standard coordinate or gradient ascent methods. These techniques are inefficient for large datasets, because we must optimize the variational parameters globally for the whole dataset. Instead, it has become common practice to use a stochastic gradient ascent (SGA) algorithm, which \cite{Hoffman13a} named ``stochastic variational inference'' (SVI). Similarly to the traditional gradient ascent, SGA updates $\lambdab$ at the $t^{th}$ iteration with
\begin{equation}\label{eqn:StochasticUpdate}
    \lambdab_{t+1} \leftarrow \lambdab_{t} + \rho_t \tilde{l}(\lambdab_t). 
\end{equation}
Here, $\tilde{l}(\lambdab)$ is a realization of the random variable $\tilde{\mathcal{L}}(\lambdab)$, so that $\mathbb{E}(\tilde{\mathcal{L}}(\lambdab)) = \nabla_{\lambdab} \mathcal{L}(\lambdab)$, and \cite{Ranganath14} showed that the gradient of ELBO with respect to the variational parameter $\lambdab$ can be written as
\begin{equation}\label{eqn:VBI:ELBOdelta}
\nabla_{\lambdab} \mathcal{L}(\lambdab) = \mathbb{E}_q\bigg[\nabla_{\lambdab} \log q(\phib|\lambdab)(\log p(\bm{d}|\phib) - \log \frac{ q(\phib|\lambdab)}{ p(\phib)})\bigg],
\end{equation}
where $\nabla_{\lambdab} \log q(\phib|\lambdab)$ is the gradient of the variational log-likelihood with respect to $\lambdab$.

SGA converges to a local maximum of $\mathcal{L}(\lambdab)$ (global for $\mathcal{L}(\lambdab)$ concave \citep{bottou-97}) when the learning rate $\rho_t$ follows the Robbins-Monro conditions \citep{robbins1951}
\begin{align}\label{eqn:VBI:RMconditions}
\sum_{t = 1}^{\infty}  \rho_t = \infty, \hspace{1cm} \sum_{t = 1}^{\infty}  \rho^2_t < \infty.
\end{align}
The bottleneck in the computation of the ELBO gradient $\nabla_{\lambdab} \mathcal{L}(\lambdab)$ is the evaluation of the log-likelihood $\log p(\bm{d}|\phib)$, which makes the traditional gradient methods as hard to scale as MCMC methods. SGA algorithms address this challenge. If we consider $N$ independent observations $d_i \sim p(d_i|\phib)$, then we can define a noisy estimate of the gradient $\nabla_{\lambdab} \mathcal{L}(\lambdab)$ as
\begin{align}\label{eqn:VBI:ELBOdeltaTilda}
\begin{split}
\tilde{\mathcal{L}}(\lambdab) &:= N \mathbb{E}_q\bigg[\nabla_{\lambdab} \log q(\phib|\lambdab)(\log p(d_I|\phib))\bigg] \\
&- \mathbb{E}_q\bigg[\nabla_{\lambdab} \log q(\phib|\lambdab)\log \frac{ q(\phib|\lambdab)}{p(\phib)}\bigg],
\end{split}
\end{align}
where $I \sim U(1,\dots, N)$ with $\mathbb{E}(\tilde{\mathcal{L}}(\lambdab)) = \nabla_{\lambdab} \mathcal{L}(\lambdab)$. Each update of $\lambdab$ computes the likelihood only for one observation $d_i$ at a time and makes the SVI scalable for large datasets. One can easily see that, under the framework for Bayesian calibration, $\mathbb{E}(\tilde{\mathcal{L}}(\lambdab)) \ne \nabla_{\lambdab} \mathcal{L}(\lambdab)$ and that the corresponding \textbf{noisy estimates are biased}.

\section{Variational Calibration of Computer Models}\label{sec:vc}
In this section, we derive the algorithm for scalable variational inference approach to Bayesian computer model calibration. The first step is finding a convenient decomposition of the likelihood $p(\bm{d}|\phib)$ that allows for an unbiased stochastic estimate of the gradient $\nabla_{\lambdab} \mathcal{L}(\lambdab)$ that depends only on a small subset of data. Multivariate copulas, and specifically their pairwise construction which we shall introduce below, provide such a decomposition. We are not the first ones to use copulas in the context of VBI. For instance, \cite{Tran2015} and \cite{Smith2020} proposed a multivariate copula as a possible variational family. However, we are the first ones using copulas in the context of computer model calibration implementing via VBI.

\subsection{Multivariate Copulas and Likelihood Decomposition}\label{subsec:vc_multivariateCopulas}
Fundamentally, a copula separates the information on the dependence structure of $N > 1$ random variables $D_1, \dots, D_N$ from their marginal distributions. Let us assume, for simplicity, that the marginal cumulative distribution functions (CDFs) $F_1, \dots, F_N$ are continuous and possess the inverse functions $F^{-1}_1, \dots, F^{-1}_N$. It follows from the probability integral transform that $U_i := F_i(D_i) \sim U(0,1)$ and conversely that $D_i = F^{-1}_i(U_i)$. With this in mind, we have
\begin{align*}
\begin{split}
&P(D_1 \le F^{-1}_1(d_1), \dots
, D_N \le F^{-1}_N(d_N)) \\
&= P(U_1 \le d_1, \dots, U_N \le d_N) := C(d_1, \dots, d_N).
\end{split}
\end{align*}
The function $C$ is a distribution with support on $[0,1]^N$, uniform marginals, and is called a copula. Under the above assumptions, a one-to-one correspondence exists between copula $C$ and the joint distribution of $\bm{D} = (D_1, \dots, D_N)^T$, as stated in the following theorem due to \cite{Skla59}. To keep the notation consistency and readability, we re-state the theorem here.
\begin{theorem}[\cite{Skla59}]\label{Sklar-thr}
	Given random variables $D_1, \dots, D_N$ with continuous marginals $F_1, \dots, F_N$ and joint distribution functions $F$, there exists a unique copula C such that for all $\bm{d} = (d_1, \dots, d_N)^T \in \mathbb{R}^N$ implies that $F(d_1, \dots, d_N) = C(F_1(d_1), \dots, F_n(d_N))$. Conversely, given $F_1, \dots, F_N$ and copula $C$, the joint distribution $F$ defined through $C(F_1(x_1), \dots, F_n(x_N))$ is an N-variate distribution functions with marginals $F_1, \dots, F_N$.
\end{theorem}
Consequently, one can write the joint probability density function (pdf) $f$ of $\bm{D}$ as 
\begin{equation}\label{eqn:VBI:copulaDensity}
f(d_1, \dots, d_N) = c(F_1(d_1), \dots, F_n(d_N)) \prod_{i = 1}^{N} f_i(d_i),
\end{equation}
where $c$ represents the copula density and $f_i$ is the marginal pdf of $D_i$.

The key reason for considering copulas is that one can decompose the $N$-dimensional copula density $c$ into a product of bivariate copulas. The starting point for this construction is a recursive decomposition of the density $f$ into a product of conditional densities
\begin{equation}\label{eqn:VBI:conditionalPdf}
f(d_1, \dots, d_N) = \prod_{i = 2}^{N} f(d_i|d_1,\dots, d_{i-1})f(d_1).
\end{equation}
For $N = 2$, the Sklar's theorem implies that 
\begin{equation}\label{eqn:VBI:f1F2:1}
    f(d_1, d_2) = c_{12}(F_1(d_1), F_2(d_2)) f_1(d_1) f_2(d_2),
\end{equation}
and
\begin{equation}\label{eqn:VBI:F1F2}
f(d_1| d_2) = c_{12}(F_1(d_1), F_2(d_2)) f_1(d_1),
\end{equation}
where
\begin{equation}\label{eqn:VBI:C1C2}
  c_{12} := c_{12}(F_1(d_1), F_2(d_2))  
\end{equation}
is a density of $C(F_1(d_1),F_2(d_2)) = F(d_1, d_2)$. Using \eqref{eqn:VBI:F1F2} for the decomposition of $(D_1, D_p)$ given $D_2, \dots, D_{p-1}$, we obtain
\begin{equation}\label{eqn:VBI:conditionalPdfCopula}
f(d_p|d_1, \dots, d_{p-1}) = (\prod_{r = 1}^{p - 2} c_{r,p;r+1, \dots, p-1}) c_{(p-1),p} \cdot f_p(d_p),
\end{equation}
where
\begin{equation}
\begin{split}
&c_{i,j;i_1, \dots, i_m} := c_{i,j;i_1, \dots, i_m}(F(d_i|d_{i_1}, \dots, d_{i_m}),F(d_j|d_{i_1}, \dots, d_{i_m}))  
 \end{split}
\end{equation}
and
\begin{equation}
\begin{split}
    &F(d_i, d_j|d_{i_1}, \dots, d_{i_m}) := C_{i,j;i_1, \dots, i_m}(F(d_i|d_{i_1}, \dots, d_{i_m}),F(d_j|d_{i_1}, \dots, d_{i_m})).
  \end{split}
\end{equation}

Using \eqref{eqn:VBI:conditionalPdf} and \eqref{eqn:VBI:conditionalPdfCopula} with the specific index choices $p = i, r = i +j$, we have that
\begin{equation}\label{eqn:VBI:Dvine}
\begin{split}
&f(d_1, \dots, d_N) = \bigg[\prod_{j = 1}^{N-1} \prod_{i = 1}^{N-j} c_{i,(i+j); (i+1), \dots, (i+j-1)}\bigg]\prod_{k = 1}^{N}f_k(d_k).
\end{split}
\end{equation}

Note that $c_{i,j;i_1, \dots, i_m}$ is a two-dimensional copula evaluated at the conditional CDFs $F(d_i|d_{i_1}, \dots, d_{i_m})$ and $F(d_j|d_{i_1}, \dots, x_{i_m})$. This decomposition is called a \textit{D-vine}. A similar class of decompositions is possible when one applies \eqref{eqn:VBI:F1F2} on $(D_{p-1}, D_p)$ given $D_1, \dots, D_{p-2}$ and sets $j = p-r, j+i = p$ to get a \textit{canonical vine (C-vine)} \citep{Kurowicka2006}:
\begin{align}\label{eqn:VBI:Cvine}
\begin{split}
&f(d_1, \dots, d_N) = f_1(d_1) \bigg[\prod_{p = 2}^{N} \prod_{k = 1}^{p-1} c_{p-k,t; 1, \dots, (p-k-1)} \cdot f_p(d_p)\bigg]= \\
&\bigg[\prod_{j = 1}^{N-1} \prod_{i = 1}^{N-j} c_{j,(j+i); 1, \dots, (j-1)}\bigg]\prod_{k = 1}^{N}f_k(d_k).
\end{split}
\end{align}

One can easily imagine that many such pair-copula decompositions exist. \cite{bedford2002} observed that these can be represented graphically as a sequence of nested trees with undirected edges, which are referred to as \textit{vine trees} and their decompositions as regular vines. Here, we focus exclusively on the D-vine and C-vine decompositions because they represent the most-studied instances of regular vines and provide an especially efficient notation. We note, however, that the following results can be extended to any regular vines.

\paragraph{Properties of vine copulas \citep{Kurowicka2006}:}
The vine copula construction is particularly attractive for two reasons. First, each pair of variables occurs only once as a conditioning set. Second, the bivariate copulas involved in the decompositions have convenient form in the case of Gaussian likelihood $f$. In particular, let $\bm{D} = (D_1, \dots, D_N)^T$ follows a multivariate normal distribution with $F_j = \Phi, j = 1, \dots, N$, where $\Phi$ is the standard normal CDF. The bivariate copula density is
\begin{equation}\label{eqn:VBI:copulaGaussDensity}
\begin{split}
&c_{i,j;i_1, \dots, i_m}(u_i, u_j) = \frac{1}{\sqrt{1 - \kappa^2}} \text{exp}\{-\frac{\kappa^2(w_i^2 + w_j^2) - 2\kappa w_i w_j}{2(1-\kappa^2)}\}.
\end{split}
\end{equation}
Here, $u_i = F(d_i|d_{i_1}, \dots, d_{i_m})$, $u_j = F(d_j|d_{i_1}, \dots, d_{i_m})$, $w_i = \Phi^{-1}(u_i)$, $w_j = \Phi^{-1}(u_j)$, and $\kappa = \rho_{i,j \cdot i_1, \dots, i_m}$ is the partial correlation of variables $i,j$ given $i_1, \dots, i_m$. The D-vine and C-vine decompositions also involve conditional CDFs, for which we need further expressions. 
Let $v$ be an index in a set $\Xi$ and $\Xi_{-v}:= \Xi \setminus v$ so that $\Xi$ contains more than one element, $F(d_j|\bm{d}_\Xi)$ is typically computed recursively as
\begin{equation}\label{eqn:VBI:Frecursive}
  F(d_j|\bm{d}_\Xi) = h(F(d_j|\bm{d}_{\Xi_{-v}}), F(d_v|\bm{d}_{\Xi_{-v}})| \rho_{j,v \cdot \Xi_{-v}})  
\end{equation}
and the function $h$ is for the Gaussian case given by
\begin{equation}\label{egn:VBI:h}
\begin{split}
&h(u_i,u_j| \rho_{i,j \cdot i_1, \dots, i_m}) = \Phi \Bigg(\frac{\Phi^{-1}(u_i) - \rho_{i,j \cdot i_1, \dots, i_m} \Phi^{-1}(u_j) }{\sqrt{1 - \rho^2_{i,j \cdot i_1, \dots, i_m}}}\Bigg).
\end{split}
\end{equation}
Lastly, the partial correlation can be also computed recursively as
\begin{equation}\label{eqn:VBI:Parcorr}
    \rho_{i,j \cdot \Xi} = \frac{\rho_{i,j \cdot \Xi_{-v}} - \rho_{i,v \cdot \Xi_{-v}}\rho_{v,j \cdot \Xi_{-v}}}{\sqrt{1- \rho^2_{i,v \cdot D_{-v}}}\sqrt{1- \rho^2_{v,j \cdot \Xi_{-v}}}}.
\end{equation}

\subsection{Scalable Algorithm with Truncated Vine Copulas}\label{subsec:vc_ScalableAlg}
We now consider the data likelihood $p(\bm{d}|\phib)$ according to \eqref{eqn:VBI:dModel} and make use of vines to construct a noisy estimate of the gradient $\nabla_{\lambdab} \mathcal{L}(\lambdab)$. We additionally assume that $N = n + s$, where $n$ is the number of observations $y_i$ from the physical process, and $s$ is the number of computer model runs $z_j$. The log-likelihood $\log p(\bm{d}|\phib)$ can be rewritten according to the D-vine decomposition as
\begin{equation}\label{egn:VBI:pDvine}
\log p(\bm{d}|\phib) = \sum_{j = 1}^{N-1} \sum_{i = 1}^{N-j} p^\Delta_{i, i+j}(\phib),   
\end{equation}
where
\begin{equation}\label{eqn:VBI:PdDefinitoin}
\begin{split}
    p^\Delta_{i, i+j}(\phib) &= \log c_{i,(i+j); (i+1), \dots, (i+j-1)} \\
    &+ \frac{1}{n-1}\big(\log p_i(d_i|\phib)
    + \log p_{i+j}(d_{i+j}|\phib)\big).
    \end{split}
\end{equation}
This can be conveniently used in the expression of the ELBO gradient. For a D-vine, we have that
\begin{equation}\label{eqn:ELBODvine}
\begin{split}
\nabla_{\lambdab} \mathcal{L}(\lambdab) &= \sum_{j = 1}^{N-1} \sum_{i = 1}^{N-j}\mathbb{E}_q\bigg[\nabla_{\lambdab} \log q(\phib|\lambdab)(p^\Delta_{i, i+j}(\phib))\bigg] \\
&- \mathbb{E}_q\bigg[\nabla_{\lambdab} \log q(\phib|\lambdab)\log\frac{q(\phib|\lambdab)}{p(\phib)}\bigg].
\end{split}
\end{equation}
The following proposition gives a noisy unbiased estimate $\tilde{\mathcal{L}}_\Delta(\lambdab)$ of the gradient \eqref{eqn:ELBODvine}. Similarly, we can derive a noisy estimate $\tilde{\mathcal{L}}_C(\lambdab)$ of the gradient using a C-vine. We leave the details of the C-vine case together with the proof of proposition \ref{propostion:Dvine} to the Appendix.

\begin{proposition}\label{propostion:Dvine}
\textit{Let $\tilde{\mathcal{L}}_\Delta(\lambdab)$ be an estimate of the ELBO gradient $\nabla_{\lambdab} \mathcal{L}(\lambdab)$ defined as}
\begin{equation*}\label{eqn:VBI:ELBODvineNoisy}
\begin{split}
\tilde{\mathcal{L}}_\delta(\lambdab) &= \frac{N(N-1)}{2}\mathbb{E}_q\bigg[\nabla_{\lambdab}\log q(\phib|\lambdab)(	p^\Delta_{I_\Delta(K)}(\phib))\bigg] \\
&- \mathbb{E}_q\bigg[\nabla_{\lambdab} \log q(\phib|\lambdab)\log \frac{ q(\phib|\lambdab)}{ p(\phib)}\bigg],
\end{split}
\end{equation*}
\textit{where $K \sim U(1, \dots, \frac{N(N-1)}{2})$, and $I_\Delta$ is the bijection}
\begin{equation*}\label{eqn:VBI:bijectionDvine}
\begin{split}
&I_\Delta:\{1, \dots, \frac{N(N-1)}{2}\} \rightarrow \{(i,i+j): i \in \{1, \dots, N-j\} \text{ for } j \in \{1, \dots N-1\}\},
\end{split}
\end{equation*}
\textit{then $\tilde{\mathcal{L}}_\Delta(\lambdab)$ is unbiased i.e., $\mathbb{E}(\tilde{\mathcal{L}}_D(\lambdab)) = \nabla_{\lambdab} \mathcal{L}(\lambdab)$}.
\end{proposition}
 
As in the case of SVI for independent data, these noisy estimates allow to update the variational parameter $\lambdab$ without the need to evaluate the whole likelihood $p(\bm{d}|\phib)$. We need to consider only the data consisting of a copula's conditioning and conditioned sets. Unfortunately, both $\tilde{\mathcal{L}}_\Delta(\lambdab)$ and $\tilde{\mathcal{L}}_C(\lambdab)$ can be relatively costly to compute for large datasets because of the recursive nature of calculations involved in the copula densities' evaluation. According to \cite{VinesTruncated01,DIMANN201352}, and \cite{BRECHMANN201519}, the most important and strongest dependencies among variables can be typically captured best by the pair copulas of the first trees. This notion motivates the use of \textit{truncated vine copulas}, where the copulas associated with the higher-order trees are set to the independence copulas. From the definition of a regular vine, one can show that the joint density $f$ can be decomposed as
\begin{equation*}\label{eqn:VBI:Rvine}
f(d_1, \dots , d_N) =  \bigg[\prod_{j = 1}^{N-1} \prod_{e \in E_i}^{} c_{j(e),k(e);\Xi(e)}\bigg]\prod_{k = 1}^{N}f_k(d_k),
\end{equation*}
where $e = j(e),k(e);\Xi(e) \in E_i$ is an edge in the $i^{th}$ tree of the vine specification. We define the truncated regular vine copula as follows.

\begin{definition}[\cite{VinesTruncated01}] 
	\textit{Let $\bm{U} = \{U_1, \dots, U_N\}$ be a random vector with uniform marginals, and let $l \in \{1, \dots, N-1\}$ be the truncation level. Let $\Pi$ denote the bivariate independence copula. Then, $\bm{U}$ is said to be distributed according to an N-dimensional l-truncated R-vine copula if $C$ is an N-dimensional R-vine copula with
	\begin{equation*}\label{eqn:VBI:IndepVine}
	C_{j(e),k(e);\Xi(e)} = \Pi \hspace{0.5cm} \forall e \in E_i \hspace{0.5cm} i = l+1, \dots, N-1.
	\end{equation*}}
\end{definition}

For the case of an \textbf{l-truncated D-vine}, we have
\begin{equation}\label{eqn:VBI:DvineTruncated}
\begin{split}
&f(d_1, \dots, d_N) = \bigg[\prod_{j = 1}^{l} \prod_{i = 1}^{N-j} c_{i,(i+j); (i+1), \dots, (i+j-1)}\bigg]\prod_{k = 1}^{N}f_k(d_k),
\end{split}
\end{equation}
and analogically to the case of D-vine with no truncation, the log-likelihood $p(\bm{d}|\phib)$ can be written as a sum of unique elements given in Proposition \ref{proposition:DvineTrunc}.
\begin{proposition}\label{proposition:DvineTrunc}
\textit{If the copula of $p(\bm{d}|\phib)$ is distributed according to an l-truncated D-vine, we can rewrite
\begin{equation}\label{eqn:DvineTruncatedP}
\log p(\bm{d}|\phib) = \sum_{j = 1}^{l} \sum_{i = 1}^{N-j} p^{\Delta_l}_{i, i+j}(\phib),
\end{equation}
where
\begin{equation}\label{eqn:VBI:pDefDvine}
\begin{split}
    p^{\Delta_l}_{i, i+j}(\phib) &=  \log c_{i,(i+j); (i+1), \dots, (i+j-1)} + \frac{1}{a_i}\log p_i(d_i|\phib) + \frac{1}{b_{i+j}} \log p_{i+j}(d_{i+j}|\phib),
    \end{split}
\end{equation}
and
\begin{align*}\label{eqn:VBI:aibiDvine}
a_i &= 2l - \bigg[(l+1-i)\mathbbm{1}_{i \le l} + (l - N +i)\mathbbm{1}_{i > N - l}\bigg], \\
b_{i+j} &= 2l -\bigg[(l+1-j-i)\mathbbm{1}_{i + j \le l} + (l - N +j+i)\mathbbm{1}_{i + j > N - l}\bigg].
\end{align*}}
\end{proposition}
The main idea for the scalable variational calibration (VC) of computer models is replacing the full log-likelihood $\log(\bm{d}|\phib)$ in the definition of ELBO with the likelihood based on a truncated vine copula. This yields the \textit{l-truncated ELBO} for the l-truncated D-vine
\begin{equation}\label{eqn:VBI:ELBODvineTruncated}
\mathcal{L}_{\Delta_l}(\lambdab) = \mathbb{E}_q\bigg[\sum_{j = 1}^{l} \sum_{i = 1}^{N-j} p^{\Delta_l}_{i, i+j}(\phib)\bigg] - KL(q(\phib|\lambdab)||p(\phib))
\end{equation}
with its gradient
\begin{equation*}\label{eqn:VBI:ELBODvineLgrad}
\begin{split}
\nabla_{\lambdab} \mathcal{L}_{\Delta_l}(\lambdab) &= \sum_{j = 1}^{l} \sum_{i = 1}^{N-j}\mathbb{E}_q\bigg[\nabla_{\lambdab} \log q(\phib|\lambdab)(p^{\Delta_l}_{i, i+j}(\phib))\bigg] - \mathbb{E}_q\bigg[\nabla_{\lambdab} \log q(\phib|\lambdab)\log\frac{q(\phib|\lambdab)}{p(\phib)}\bigg].
\end{split}
\end{equation*}
The following proposition gives a noisy unbiased estimate $\tilde{\mathcal{L}}_{\Delta_l}(\lambdab)$ of the gradient $\nabla_{\lambdab} \mathcal{L}_{\Delta_l}(\lambdab)$. We can analogously derive an unbiased estimate $\tilde{\mathcal{L}}_{C_l}(\lambdab)$ of the gradient using C-vine (see the Appendix for details and the proof).

\begin{proposition}\label{propostion:DvineTruncUnbiase}
\textit{Let $\tilde{\mathcal{L}}_{\Delta_l}(\lambdab)$ be an estimate of the ELBO gradient $\nabla_{\lambdab} \mathcal{L}_{\Delta_l}(\lambdab)$ defined as}
\begin{equation*}\label{eqn:VBI:ELBODvineTruncNoisy}
\begin{split}
&\tilde{\mathcal{L}}_{\Delta_l}(\lambdab) = \frac{l(2N-(l + 1))}{2}\mathbb{E}_q\bigg[\nabla_{\lambdab} \log q(\phib|\lambdab)(	p^{\Delta_l}_{I_{\Delta_l}(K)}(\phib))\bigg] -\mathbb{E}_q\bigg[\nabla_{\lambdab} \log q(\phi|\lambdab)\log\frac{ q(\phi|\lambdab)}{ p(\phi)}\bigg],
\end{split}
\end{equation*}
\textit{where $K \sim U(1, \dots, \frac{l(2N-(l + 1))}{2})$, and $I_{\Delta_l}$ is the bijection}
\begin{equation*}\label{eqn:VBI:bijectionDvineTrunc}
\begin{split}
&I_{\Delta_l}:\{1, \dots, \frac{l(2N-(l + 1))}{2}\} \rightarrow \{(i,i+j): i \in \{1, \dots, N-j\} \text{ for } j \in \{1, \dots l\}\},
\end{split}
\end{equation*}
\textit{then $\tilde{\mathcal{L}}_{\Delta_l}(\lambdab)$ is unbiased i.e., $\mathbb{E}(\tilde{\mathcal{L}}_{\Delta_l}(\lambdab)) = \nabla_{\lambdab} \mathcal{L}_{\Delta_l}(\lambdab)$}.
\end{proposition} 

Considering the l-truncated ELBO \eqref{eqn:VBI:ELBODvineTruncated}, our proposed algorithm for the VC of computer models with truncated vine copulas is stated in the Algorithm~\ref{alg-truncated}. Note that $\tilde{\mathcal{L}}_{\Delta_l}(\lambda)$ does not have closed form expression in general due to expectations involved in the computation. Therefore, we resort to a Monte Carlo (MC) approximation of $\tilde{\mathcal{L}}_{\Delta_l}(\lambda)$ using samples from the variational distribution.
\begin{algorithm}[h]
\DontPrintSemicolon
	\caption{Variational calibration with truncated D-vine copulas.\label{alg-truncated}}
		\addtocontents{loa}{\vskip 12pt}
		\KwIn{Data $\bm{d}$, mean and covariance functions for GPs, variational family $q(\phib|\lambdab)$, \textbf{truncation level l}}
		$\lambda \leftarrow$ random initial value\;
		$t \leftarrow 1$\;
		\Repeat{change of $\lambdab$ is less than $\epsilon$}{
		\For{$s = 1$ to $S$}{
		$\phib[s] \sim q(\phib|\lambdab)$
		}
		$K \leftarrow U(1, \dots, \frac{l(2N-(l + 1))}{2})$\;
		$\rho \leftarrow$ $t^{\textrm{th}}$ value of a Robbins-Monro sequence\;
		$\lambdab \leftarrow \lambdab + \rho \frac{1}{S}\sum_{s=1}^{S} \bigg[\frac{l(2N-(l + 1))}{2} \nabla_{\lambdab} \log q(\phib[s]|\lambdab) \times \big(	p^{\Delta_l}_{I_{\Delta_l}(K)}(\phib[s]) -\frac{2}{l(2N-(l + 1))} \log\frac{ q(\phib[s]|\lambdab)}{p(\phib[s])}\big)\bigg]$\;
		$t \leftarrow t + 1$\;}
\end{algorithm}

\paragraph{Scalability Discussion:} The complexity of bivariate copula evaluation depends on the size of conditioning dataset due to the recursive nature of the calculations \citep{Kurowicka2006}. From the vine tree construction, the cardinality of the conditioning set for D-vine and C-vine is in the worst case $N-2$. Nevertheless, on average, we can do better. Indeed, let $X$ be the cardinality of the conditioning set in $p^\Delta_{I_\Delta(K)}$ (or $p^C_{I_C(K)}$),
then
\begin{equation}\label{eqn:avg_vine}
P(X = i) = \frac{N - (i + 1)}{\binom{n}{2}} \qquad  \text{for} \; i \in \{0, \dots, N-2\}
\end{equation}
and $\mathbb{E}(X) = \frac{N-2}{3}$. The cardinality of conditioning set is on average roughly $N/3$. On the other hand, the cardinality of conditioning set is for the case of Algorithm \ref{alg-truncated} at most $l - 1$. Now, let $X_l$ be the cardinality of the conditioning set in the updating step of the variational parameter $\lambda$ in the Algorithm \ref{alg-truncated}, then
\begin{equation}\label{eqn:avg_vine_truncated}
P(X_l = i) = \frac{N - (i + 1)}{\frac{l(2N-(l + 1))}{2}} \qquad \text{for} \; i \in \{0, \dots, l-1\},
\end{equation}
and $\mathbb{E}(X_l)  = [(l-1)(3N -2l -2)]/[3(2N - l - 1)]$. $\mathbb{E}(X_l) \approx 2$ for $N = 10^5$ and truncation level $l = 5$, which is a significant improvement to the average case $p^\Delta_{I_\Delta(K)}$ and $p^C_{I_C(K)}$ ($\approx 33333$ for $N = 10^5$). This provides a heuristic yet compelling argument for the scalability.

\section{Implementation Details}\label{sec:implemenation}
\subsection{Selection of Truncation Level}\label{subsec:implementation_truncation}
Selection of the truncation level $l$ is an important element in effective approximation of the posterior distribution $p(\phib|\bm{d})$ under Algorithm~\ref{alg-truncated}. \cite{DIMANN201352} propose a sequential approach for selection of $l$ in the case of vine estimation. One sequentially fits models with an increasing truncation level until the quality of fit stays stable or computational resources are depleted. We adopt similar idea for the case of VC of computer models with vine copulas. Let $\lambdab(l)$ represents the value of variational parameter estimated with Algorithm~\ref{alg-truncated} for a fixed truncation level $l$. One can then sequentially increase $l$ until $\|\lambdab(l+1) - \lambdab(l)\| < \epsilon$ for some norm $\| \cdot \|$ and a desired tolerance $\epsilon$.

\subsection{Variance Reduction of Monte Carlo Approximations}\label{subsec:implementaion_variance}
The computational convenience of MC approximations of the gradient estimators based on the l-truncated D-vine and C-vine copulas $\tilde{\mathcal{L}}_{\Delta_l}(\lambdab)$ and $\tilde{\mathcal{L}}_{C_l}(\lambdab)$ (see Section \ref{subsec:vc_ScalableAlg}) is typically accompanied by their large variance. The consequence in practice is the need for small step size $\rho_t$ in the SGA portion of Algorithm~\ref{alg-truncated} which results in a slower convergence. In order to reduce the variance of MC approximations, we adopt similar approach as \cite{OBBVI} and use Rao-Blackwellization \citep{RB}, control variates (CV) \citep{Ross:2006}, and importance sampling. The reminder of this section focuses on the case of D-vine decomposition, see Appendix for the derivations for C-vines.

\subsubsection{Rao-Blackwellization}\label{subsec:implementation_rb}
The idea of Rao-Blackwellization is to replace the noisy estimate of gradient with its conditional expectation with respect to a subset of $\phib$. For simplicity, let us consider a situation with $\phib = (\phi_1, \phi_2) \in \mathbb{R}^2$ and variational family $q(\phib|\lambdab)$ that factorizes into $q(\phi_1|\lambdab_1)q(\phi_2|\lambdab_2)$. Additionally, let $\hat{\mathcal{L}}_{\lambdab}(\phi_1, \phi_2)$ be the MC approximation of the gradient $\nabla_{\lambdab} \mathcal{L}(\lambdab)$. Now, the conditional expectation $\mathbb{E}[\hat{\mathcal{L}}_{\lambdab}(\phi_1, \phi_2)| \phi_1]$ is also an unbiased estimate of $\nabla_{\lambdab} \mathcal{L}(\lambda)$ since $\mathbb{E}_q(\mathbb{E}[\hat{\mathcal{L}}_{\lambdab}(\phi_1, \phi_2)| \phi_1]) = \mathbb{E}_q(\hat{\mathcal{L}}_{\lambdab}(\phi_1, \phi_2))$ and
\begin{equation*}
\begin{split}
&\mathbb{V}ar_q(\mathbb{E}[\hat{\mathcal{L}}_{\lambdab}(\phi_1, \phi_2)| \phi_1]) = \mathbb{V}ar_q(\hat{\mathcal{L}}_{\lambdab}(\phi_1, \phi_2)) - \mathbb{E}[(\hat{\mathcal{L}}_{\lambdab}(\phi_1, \phi_2) - \mathbb{E}[\hat{\mathcal{L}}_{\lambdab}(\phi_1, \phi_2)| \phi_1])^2]
\end{split}
\end{equation*}
shows that $\mathbb{V}ar_q(\mathbb{E}[\hat{\mathcal{L}}_{\lambdab}(\phi_1, \phi_2)| \phi_1]) \le \mathbb{V}ar_q(\hat{\mathcal{L}}_{\lambdab}(\phi_1, \phi_2))$. The factorization of the variational family also makes the conditional expectation straightforward to compute as
\begin{equation*}
\begin{split}
&\mathbb{E}[\hat{\mathcal{L}}_{\lambdab}(\phi_1, \phi_2)| \phi_1] = \int_{\phi_2} \mathbb{E}[\hat{\mathcal{L}}_{\lambdab}(\phi_1, \phi_2)] \frac{q(\phi_1|\lambdab_1)q(\phi_2|\lambdab_2)}{q(\phi_1|\lambdab_1)} \diff \phi_2 = \mathbb{E}_{q(\phi_2|\lambdab_2)}(\hat{\mathcal{L}}_{\lambdab}(\phi_1, \phi_2)),
\end{split}
\end{equation*}
i.e., we just need to integrate out some variables. Let us consider the MC approximation of the gradient estimator $\tilde{\mathcal{L}}_{\Delta_l}(\lambdab)$. The $j^{th}$ entry of the Rao-Blackwellized estimator is
\begin{equation*}
\begin{split}
\frac{1}{S}\sum_{s=1}^{S}&\bigg[\frac{l(2N-(l + 1))}{2}\nabla_{\lambdab_j} \log q(\phi_j[s]|\lambdab_j)\big(\tilde{p}_{(j)}(\phib[s]) - \frac{2}{l(2N-(l + 1))}\log\frac{ q(\phi_j[s]|\lambdab_j)}{p(\phi_j[s])}\big)\bigg],
\end{split}
\end{equation*}
where $\tilde{p}_{(j)}(\phib)$ are the components of $p^{\Delta_l}_{I_{\Delta_l}(K)}(\phib)$ that include $\phi_j$.
\subsubsection{Control Variates}\label{subsec:implementation_cv}
To further reduce the variance of the MC approximations we will replace the Rao-Blackwellized estimate above with a function that has the same expectation but again smaller variance. For illustration, let us first consider a target function $\xi(\phib)$ whose variance we want to reduce, and a function $\psi(\phib)$ with finite expectation. Define
\begin{equation}\label{eqn:VBI:CV}
\hat{\xi}(\phib) = \xi(\phib) - a(\psi(\phib) - \mathbb{E}_q[\psi(\phib)]),
\end{equation}
where $a$ is a scalar and $\mathbb{E}_q(\hat{\xi}(\phib)) = \mathbb{E}_g[\xi(\phib)]$. The variance of $\hat{\xi}(\phib)$ is
\begin{equation}\label{eqn:VBI:CVvar}
\begin{split}
\mathbb{V}ar_q(\hat{\xi}(\phib)) &= \mathbb{V}ar_q(\xi(\phib)) + a^2\mathbb{V}ar_q(\psi(\phib)) -2a\mathbb{C}ov_q(\xi(\phib),\psi(\phib)).
\end{split}
\end{equation}
This shows that a good choice for function $\psi(\phib)$ is one that has high covariance with $\xi(\phib)$. Moreover, the value of $a$ that minimizes \eqref{eqn:VBI:CVvar} is
\begin{equation}\label{eqn:VBI:CVa}
a^*= \frac{\mathbb{C}ov_q(\xi(\phib),\psi(\phib))}{\mathbb{V}ar_q(\psi(\phib))}.
\end{equation}

Let us place the CV back into the context of calibration. Meeting the above described criteria, \cite{Ranganath14} propose $\psi(\phib)$ to be $\nabla_{\lambdab} \log q(\phib|\lambdab)$, because it depends only on the variational distribution and has expectation zero. We can now set the target function $\xi(\phib)$ to be \begin{equation*}
\begin{split}
\frac{l(2N-(l + 1))}{2}\nabla_{\lambdab_j} &\log q(\phi_j|\lambdab_j)\bigg(\tilde{p}_{(j)}(\phib) - \frac{2}{l(2N-(l + 1))}\log\frac{ q(\phi_j|\lambdab_j)}{p(\phi_j)}\bigg),
 \end{split}
\end{equation*}
which gives the following $j^{th}$ entry of the MC approximation of the gradient estimator $\tilde{\mathcal{L}}_{\Delta_l}(\lambdab)$ with CV
\begin{align*}\label{eqn:VBI:CVELBO}
&\tilde{\mathcal{L}}_{\Delta_l}^{CV(j)}(\lambdab) =  \frac{1}{S}\sum_{s=1}^{S}\bigg[\frac{l(2N-(l + 1))}{2}\nabla_{\lambdab_j} \log q(\phi_j[s]|\lambdab_j)\big(\tilde{p}_{(j)}(\phib[s])- \frac{2 (\log\frac{ q(\phi_j[s]|\lambdab_j)}{p(\phi_j[s])} + \hat{a}^\Delta_j)}{l(2N-(l + 1))} \big)\bigg],
\end{align*}
where $\hat{a}^{\Delta}_j$ is the estimate of $a^*$ based on additional independent draws from the variational approximation (otherwise the estimator would be biased).

\subsubsection{Importance sampling}\label{subsec:implementation_importanceSampling}
The ultimate variance reduction technique used is the importance sampling. We refer to \cite{OBBVI} for full description of the method and illustration of its efficiency in the VBI framework. Fundamentally, instead of taking samples from the variational family $q(\phib|\lambdab)$ to carry out the MC approximation of the ELBO gradient estimate, we will take samples from an overdispersed distribution $r(\phib|\lambdab, \tau)$ in the same family that depends on an additional dispersion parameter $\tau > 1$. Namely, we can write the estimate $\tilde{\mathcal{L}}_{\Delta_l}(\lambda)$ as 
\begin{equation*}
\begin{split}
\mathbb{E}_{r(\phib|\lambdab, \tau)}&\bigg[\frac{l(2N-(l + 1))}{2}\nabla_{\lambdab} \log q(\phib|\lambdab)(	p^{\Delta_l}_{I_{\Delta_l}(K)}(\phib) - \frac{2}{l(2N-(l + 1))} \log\frac{ q(\phib|\lambdab)}{p(\phib)})w(\phib)\bigg],
\end{split}
\end{equation*}
where $w(\phib) = q(\phib|\lambdab) / r(\phib|\lambdab, \tau)$ is the importance weight which guarantees the estimator to be unbiased. The reason to formulate the $\tilde{\mathcal{L}}_{\Delta_l}(\lambda)$ this way comes from the fact the optimal proposal \citep{robert2005monte} distribution to form the MC estimate is not $q(\phib|\lambdab)$, but rather 
\begin{equation}\label{eqn:VBI:optimalProposal}
    r^*(\phib) \propto q(\phib|\lambdab) |\xi(\phib)|,
\end{equation}
where
\begin{equation}\label{eqn:VBI:IsamplingXi}
\begin{split}
&\xi(\phib) = \frac{l(2N-(l + 1))}{2}\nabla_{\lambdab} \log q(\phib|\lambdab)\bigg(	p^{\Delta_l}_{I_{\Delta_l}(K)}(\phib) - \frac{2}{l(2N-(l + 1))} \log\frac{ q(\phib|\lambdab)}{p(\phib)}\bigg).
\end{split}
\end{equation}
However, the normalizing constant for the optimal $r^*(\phib)$ is intractable, and so \cite{OBBVI} propose that an overdispersed version of the variational family that assigns higher probability to the tails of $q(\phib|\lambdab)$ is closer to the optimum than $q(\phib| \lambdab)$ itself. For example, if the value of $\lambdab$ makes the variational family a poor fit, then the samples $\phib[s] \sim q(\phib|\lambdab)$ have a high value for the variational distribution but low for the true posterior. On the other hand, $r^*(\phib)$ proposes values of $\phib[s]$ for which $\xi(\phib)$ is large that are in the tails of $p(\phib|\lambdab)$.

To see how the importance sampling leads to the reduction of variance of the MC estimates, let us consider the following estimator
\begin{equation}\label{eqn:VBI:MCsimple}
    \widehat{\mathcal{L}}_{MC} = \frac{1}{S} \sum_{s = 1}^S \xi(\phib[s]), \qquad \phib[s] \sim p(\phib|\lambdab),
\end{equation}
then
\begin{equation}\label{eqn:VBI:MCsimpleVar}
    \mathbb{V}ar\big[\widehat{\mathcal{L}}_{MC}\big] = \frac{1}{S} \mathbb{E}_q\big[ \xi^2(\phib)\big] - \frac{1}{S}\big[\tilde{\mathcal{L}}_{\Delta_l}(\lambda)\big]^2.
\end{equation}
Similarly, we can derived the variance of the MC estimator with the importance weights
\begin{equation}\label{eqn:VBI:MCsimpleIm}
    \widehat{\mathcal{L}}^O_{MC} = \frac{1}{S} \sum_{s = 1}^S \xi(\phib[s])\frac{q(\phib[s]|\lambdab)}{r(\phib[s]|\lambdab, \tau)}, \quad \phib[s] \sim r(\phib|\lambdab, \tau),
\end{equation}
as
\begin{equation}\label{eqn:VBI:MCsimpleVarIm}
    \mathbb{V}ar\big[\widehat{\mathcal{L}}^O_{MC}\big] = \frac{1}{S} \mathbb{E}_q\big[ \xi^2(\phib) \frac{q(\phib|\lambdab)}{r(\phib|\lambdab, \tau)}\big] - \frac{1}{S}\big[\tilde{\mathcal{L}}_{\Delta_l}(\lambda)\big]^2.
\end{equation}
Now, if we choose the distribution $r(\phib|\lambdab, \tau)$ such that
\begin{equation}\label{eqn:VBI:imp:cond}
    \mathbb{E}_q\big[ \xi^2(\phib) \frac{q(\phib|\lambdab)}{r(\phib|\lambdab, \tau)}\big] \le \mathbb{E}_q\big[ \xi^2(\phib)\big],
\end{equation}
the variance reduction will be achieved. The optimal $r^*$ obviously satisfies the condition \eqref{eqn:VBI:imp:cond}. \cite{OBBVI} show that the choice of overdispersed version of the variational family $q(\phib|\lambdab)$ has similar effect on the variance reduction as the optimal $r^*$. The details on the form of overdispersed families for specific variational families are discussed later in Section \ref{subsec:implementation_Param}.

Combining the ideas of the Rao-Blackwellization, CV, and importance sampling, we have the following $j^{th}$ entry of the MC approximation of the gradient estimator $\tilde{\mathcal{L}}_{\Delta_l}(\lambdab)$ 
\begin{align*}\label{eqn:VBI:CVisELBO}
\begin{split}
&\tilde{\mathcal{L}}_{\Delta_l}^{OCV(j)}(\lambdab) =\\
&\sum_{s=1}^{S}\bigg[\frac{l(2N-(l + 1))}{2S}\nabla_{\lambdab_j} \log q(\phi_j[s]|\lambdab_j)(\tilde{p}_{(j)}(\phi[s])-\\
&\qquad \frac{2(\log\frac{ q(\phi_j[s]|\lambdab_j)}{p(\phi_j[s])} + \tilde{a}^\Delta_j)}{l(2N-(l + 1))})w(\phi_j[s])\bigg],
\end{split}
\end{align*}
where $\phib[s] \sim r(\phib|\lambdab, \tau)$ and 

\begin{strip}
\begin{equation}
\qquad \quad \tilde{a}^\Delta_j = \frac{\widehat{\mathbb{C}ov}_r(\frac{l(2N-(l + 1)) w(\phi_j)}{2}\nabla_{\lambdab_j} \log q(\phi_j|\lambdab_j)(	\tilde{p}_{(j)}(\phib) -\frac{2\log\frac{ q(\phi_j|\lambdab_j)}{p(\phi_j)}}{l(2N-(l + 1))}), \nabla_{\lambdab_j} \log q(\phi_j|\lambdab_j) w(\phi_j))}{\widehat{\mathbb{V}ar}_r(\nabla_{\lambdab_j} \log q(\phi_j|\lambdab_j) w(\phi_j))}.
\end{equation}
\end{strip}

The extension of the Algorithm~\ref{alg-truncated} with the variance reductions of the MC approximations due to Rao-Blackwellization, CV, and importance sampling is summarized in the Algorithm~\ref{alg:TruncatedDfinal}.
\begin{algorithm}[h]
\DontPrintSemicolon
	\caption{Variational calibration with truncated D-vine copulas II.\label{alg:TruncatedDfinal}}
	\addtocontents{loa}{\vskip 12pt}
		\KwIn{Data $\bm{d}$, mean and covariance functions for GPs, variational family $q(\phib|\lambdab)$, dispersion parameter $\tau$, \textbf{truncation level l}}
		$\lambda \leftarrow$ random initial value\;
		$t \leftarrow 1$\;
		\Repeat{change of $\lambdab$ is less than $\epsilon$}{
		\For{$s = 1$ to $S$}{
		$\phib[s] \sim r(\phib|\lambdab, \tau)$
		}
		$K \leftarrow U(1, \dots, \frac{l(2N-(l + 1))}{2})$\;
		$\rho \leftarrow$ $t^{\textrm{th}}$ value of a Robbins-Monro sequence\;
		$
		\lambdab \leftarrow \lambdab + \bm{\rho}  \sum_{s=1}^{S}\bigg[\frac{l(2N-(l + 1))}{2S}\nabla_{\lambdab_j} \log q(\phi_j[s]|\lambdab_j) \times \big(\tilde{p}_{(j)}(\phib[s])-\frac{2(\log\frac{ q(\phi_j[s]|\lambdab_j)}{p(\phi_j[s])} + \tilde{a}^\Delta_j)}{l(2N-(l + 1))}\big)w(\phi_j[s])\bigg]
		$\;
		$t \leftarrow t + 1$\;}
\end{algorithm}

\subsection{Choice of the learning rate}\label{subsec:implementation_AdaGrad}
Even though the SGA is straightforward in its general definition, the choice of learning rate $\rho_t$ can be challenging in practice. Ideally, one would want the rate to be small in the situations where the noisy estimates of the gradient have large variance and vice-versa. The elements of variational parameter $\lambdab$ can also differ in scale, and one needs to set the learning rate so that the SGA can accommodate even the smallest scales. The rapidly increasing usage of machine learning techniques in recent years produced various algorithms for element-wise adaptive-scale learning rates. We use the adaptive gradient (AdaGrad) algorithm \citep{AdaGrad} which has been considered in similar problems before, e.g., \cite{Ranganath14}, however, there are other popular algorithms such as the ADADELTA \citep{Zeiler2012} or the RMSProp \citep{Tieleman2012}. Let $\gb_T$ be the gradient used in the $T^{th}$ step of the SGA algorithm, and $\Gb_t$ be the matrix consisting of the sum of the outer products of these gradients across the first $t$ iterations, namely
\begin{equation}\label{eqn:VBI:Gt}
\Gb_t = \sum_{T = 1}^{t} \gb_T\gb_T^T.
\end{equation}
The AdaGrad defines the element-wise adaptive scale learning rate as $\bm{\rho}_t = \eta \cdot \text{diag}(\Gb_t)^{-1/2}$, where $\eta$ is the initial learning rate. It is a common practice, however, to add a small constant value to $\text{diag}(\Gb_t)$ (typically of order $10^{-6}$) to avoid division by zero.

\subsection{Parametrizations}\label{subsec:implementation_Param}
\paragraph{Variational families.} We use a Gaussian distribution for real valued components of $\phib$ and a gamma distribution for positive variables. Both families are parametrized in terms of their mean and standard deviation. Moreover, in order to avoid constrained optimization, we transform all the positive variational parameters $\lambdab$ to $\tilde{\lambdab} =\log{(e^{\lambdab} - 1)}$ and optimize with respect to $\tilde{\lambdab}$.
\paragraph{Overdispersed families.} Given a fixed dispersion coefficient $\tau$, the overdispersed Gaussian distribution with mean $\mu$ and standard deviation $\sigma$ is a Gaussian distribution with mean $\mu$ and standard deviation $\sigma \sqrt{\tau}$. The overdispersed gamma distribution with mean $\mu$ and standard deviation $\sigma$ is a gamma distribution with mean $\mu + (\tau - 1) \frac{\sigma^2}{\mu}$ and standard deviation $\frac{\sigma \sqrt{\tau\mu^2 +\tau\sigma^2(\tau - 1)}}{\mu}$ \citep{OBBVI}.

\section{Applications}\label{sec:applications}
\subsection{Simulation study}\label{subsec:application_simulation}
In this section, we study Algorithm \ref{alg:TruncatedDfinal} in a simulated scenario, where we first demonstrate the method's fidelity in approximating the posterior distribution of calibration parameters $p(\thetab|\bm{d})$ and substantiate the indispensability of the variance reduction techniques described in Section \ref{sec:implemenation} in order to achieve convergence. Second, we show the scalability of our method in comparison to the popular MH algorithm and the NUTS.

\setlength{\tabcolsep}{3pt}
\renewcommand{\arraystretch}{0.7}
\small
\begin{table}[h]
\centering
\begin{tabular}{l|c|c}
		\hline \hline \noalign{\smallskip}
		& GP mean & GP covariance function \\
		\noalign{\smallskip} \hline \noalign{\smallskip}
		$f_m$ & $\theta_1 cos(t_1) + \theta_2 sin(t_2)$ & $\eta_f \cdot \text{exp}(-\frac{||\tb - \tb' ||^2}{2l^2_t} - \frac{||\thetab - \thetab' ||^2}{2l^2_\theta})$  \\
		$\delta$ & $\beta_\delta$ & $\eta_\delta \cdot \text{exp}(-\frac{||\tb - \tb' ||^2}{2l^2_\delta})$  \\
 \noalign{\smallskip} \hline \hline \noalign{\smallskip}
\end{tabular}
\caption{The specification of GPs for the simulation study.\label{table:VBI:GPS}}
\end{table}
\normalsize

Let us consider a simple scenario following the model \eqref{eqn:completeModel} with a two-dimensional calibration parameter $\thetab = (0.39, 0.60)$ that was obtained as a sample from its prior distribution $p(\thetab)$ and a two-dimensional input variable $\tb = (t_1,t_2)$. We model $f_m(\tb,\thetab)$ and $\delta(\tb)$ with GPs according to the specifications in Table \ref{table:VBI:GPS} with the particular choices of $\eta_f = \frac{1}{30}$, $l_t = 1$, $l_\theta = 1$, $\eta_\delta = \frac{1}{30}$, $l_\delta = \frac{1}{2}$, and $\beta_\delta = 0.15$. 

We choose the variational family to be the mean-field family with Gaussian distributions for real valued parameters and gamma distributions for positive variables following the parametrization discussed in Section \ref{subsec:implementation_Param}. The variational parameters are initialized to match the prior distributions, and we use the AdaGrad for the learning rate updates.

\subsubsection{Calibration}\label{subsubsec:application_simulation_calib}

For the purpose of model calibration, we sampled the data $\bm{d}$ jointly from the prior with the experimental noise following $\cN(0, \frac{1}{100})$.  The calibration parameter values for the model runs $\zb$ were selected on a uniform grid over $[ 0, 1 ]^2$ and the inputs $\tb$ over $[ 0, 3 ]^2$. For the first set of experiments, the size of the dataset was $N = 225$ with $n = 144$ and $s =81$. We used 50 samples from the variational family to approximate the expectations in Algorithm \ref{alg:TruncatedDfinal} and 10 samples to implement the control variates.

\begin{figure*}[h!]
	\begin{centering}
		\includegraphics[width=1\textwidth]{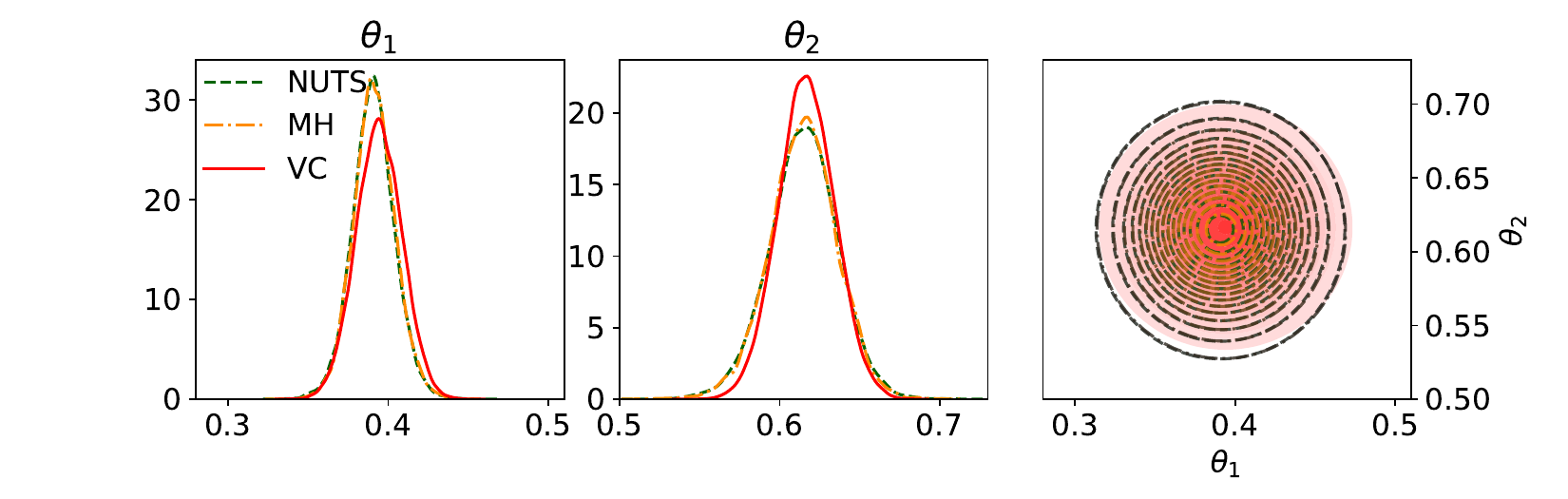}
		\caption{The approximate posterior distributions for the target calibration parameters. The VC (Algorithm \ref{alg:TruncatedDfinal}) was carried out using $l = 3$ truncated D-vine and compared with the results from the NUTS and the MH algorithm.\label{fig:VBI:applicationCalibration}}
		\label{fig:Sim1:1:Params}
	\end{centering}
\end{figure*}

Figure \ref{fig:VBI:applicationCalibration}  demonstrates the quality of the variational approximation (Algorithm \ref{alg:TruncatedDfinal}) in comparison to the MH algorithm and the NUTS. We can see that our method was able to accurately match both MCMC-based approximations with a minor deviation in $\theta_1$. It is important to note, however, that the variance reduction through the combination of the Rao-Blackwellization, control variates, and importance sampling was necessary to achieve meaningful convergence.

\begin{figure*}[h!]
	\begin{centering}
		\includegraphics[width=1\textwidth]{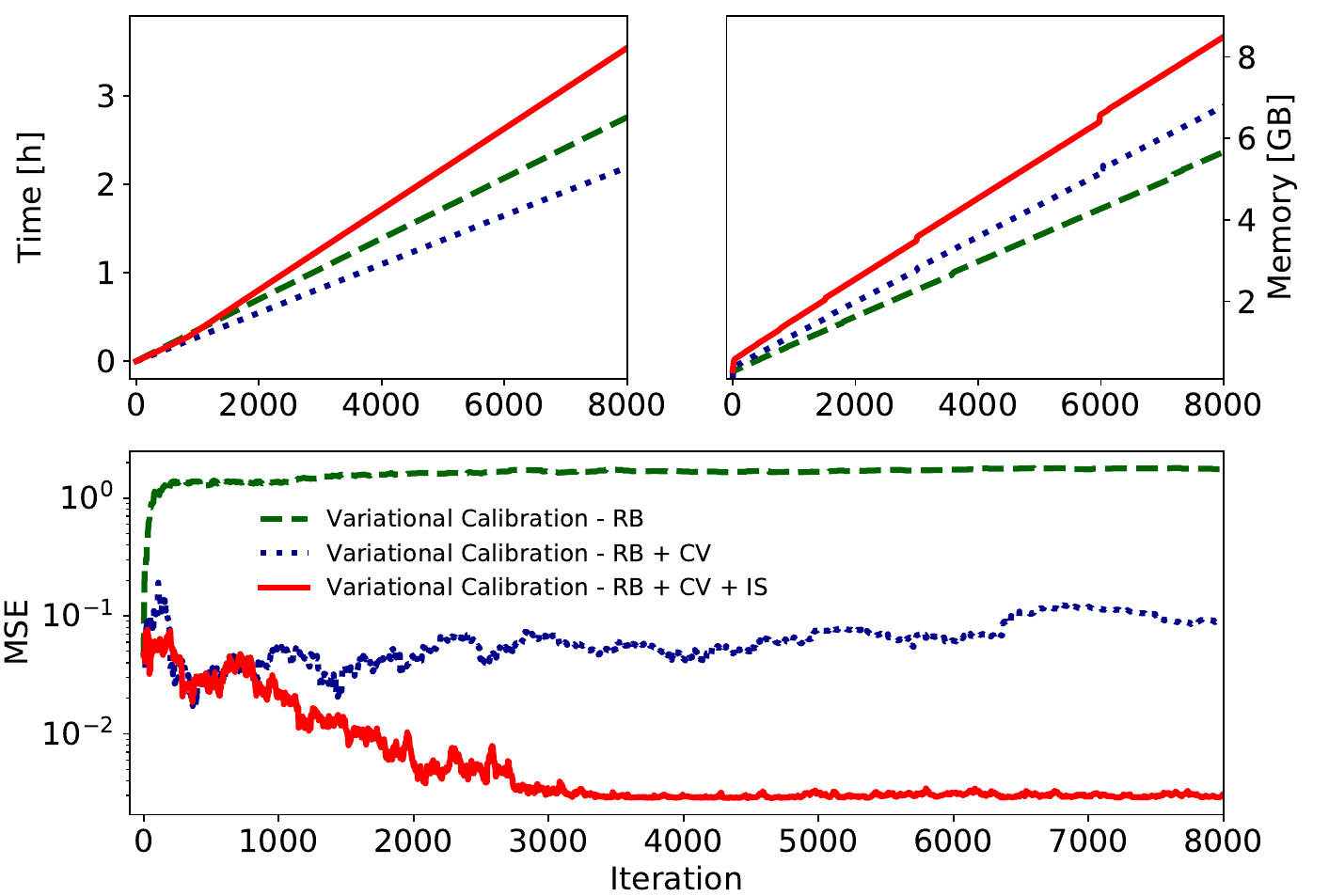}
		\caption{The evolution of MSE of the posterior predictive means based on the VC with cumulatively implemented variance reduction techniques described in Section \ref{subsec:implementaion_variance}. The figure is based on an independently generated set of 50 testing points. Time and memory demands for each of the implementations are also plotted the VC (Algorithm \ref{alg:TruncatedDfinal})  was carried out using $l = 3$ truncated D-vine. \label{fig:VBI:MSEsimulation}}
	\end{centering}
\end{figure*}

In particular, Figure \ref{fig:VBI:MSEsimulation} shows the mean squared error (MSE) of the posterior predictive means, evaluated on an independently generated set of 50 data points, based on the VC with cumulatively implemented variance reduction techniques. Algorithm \ref{alg:TruncatedDfinal} which employs the importance sampling clearly outperforms the calibration with only the Rao-Blackwellization and the calibration with control variates. In fact, each additional attempt to reduce the variance tends to decrease the MSE by one order of magnitude. There is naturally a time and space (memory) cost associated with each variance reduction technique. Figure \ref{fig:VBI:MSEsimulation} shows that the control variates and the importance sampling practically double the time per iteration of the algorithm. This additional complexity is, however, outweighed by the gain in the MSE reduction. The increase in memory consumption is less significant and is due to the storage of dispersion coefficients used for importance sampling and samples needed to compute control variates. Note that the memory consumed by the algorithms rises over time, because we chose to store the values of variational parameters during each step; the memory demands can be dramatically reduced if we drop these intermediate results. 

For completeness, in Table \ref{tab:mse}, we also compare the MSE of MCMC approximations and the VC at the point of convergence of the algorithms. The resulting errors in the predictions were, for all the practical purposes, equivalent.

\begin{table}[h!]
	\caption{Comparison of the MSE for the simple scenario using the MH, the NUTS, and the VC algorithms. \label{tab:mse}}
	\begin{tabular}{l|l} 
		\hline \hline
		Algorithm & $MSE$ \\ \hline 
		VC with RB + CV + IS & $2.9 \times 10 ^ {-3}$ \\
		Metropolis-Hastings & $3.0 \times 10 ^ {-3}$  \\
		No-U-Turn & $3.0 \times 10 ^ {-3}$ \\ \hline \hline
	\end{tabular}
\end{table}

\subsubsection{Scalability}\label{subsubsec:application_simulation_scal}

We now significantly increase the size of the dataset from $N=225$ to $0.5 \times 10^4$ and eventually to $2\times 10^4$ with the simulated experimental measurements and the model runs split equally ($n = s$). For better numerical stability, we expand the space of the input variables to $\tb \in [0,10] ^ 2$ and select those using the Latin hypercube design. We also enlarge the testing dataset to 200 points. All the remaining simulation parameters are unchanged. The conventional MCMC methods are already impractical for the purpose of Bayesian calibration with these moderately large amounts of data. We were able to obtain only around 600 posterior samples in the case of $N = 1 \times 10^4$ and about 120 for $N = 2 \times 10^4$ in $25$ hours of sampling using the MH algorithm (significantly less with the NUTS).

Algorithm \ref{alg:TruncatedDfinal} (D-vine with truncation $l=5$) converges to the predictive MSE of about 0.003 under 4 hours for $N = 2 \times 10^4$ and 2 hours for $N = 0.5 \times 10^4$. It took similar time for the MH to achieve this MSE value for $N = 0.5 \times 10^4$ but almost 25 hours for the NUTS. Once we increased the data size to $2 \times 10^4$, neither the NUTS nor the MH were able to achieve a similar predictive MSE as the VC within the 25 hour window allotted for sampling. In fact, they were by an order of magnitude larger. It is important to mention that both MCMC-based algorithms have also substantially larger memory demands than the VC. See Appendix \ref{app:mem} for more details.

\begin{figure}[h!] 
	\begin{centering}
		\includegraphics[width=0.8\textwidth]{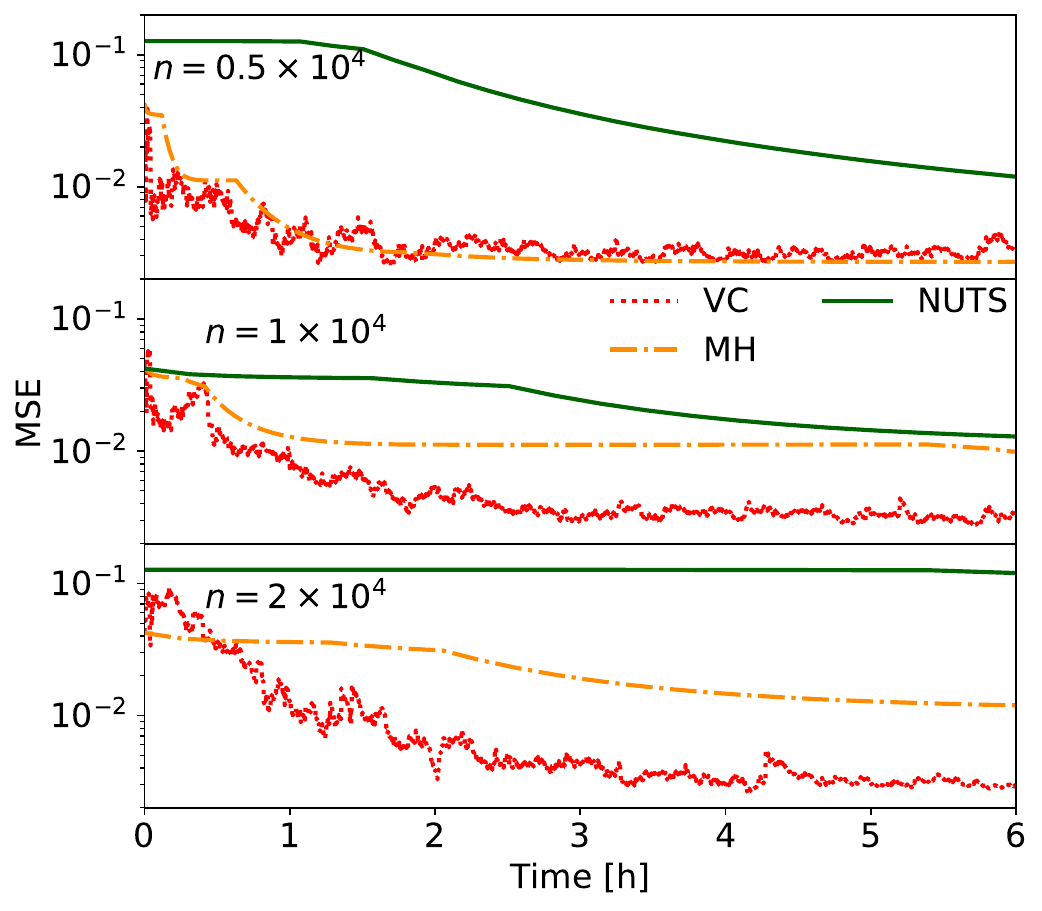}
		\caption{The evolution of the MSE of the posterior predictive means based on the VC (Algorithm \ref{alg:TruncatedDfinal}), the MH algorithm, and the NUTS. The figure is based on an independently generated set of 200 testing points. The VC (Algorithm \ref{alg:TruncatedDfinal}) was carried out using $l = 5$ truncated D-vine.}
		\label{fig:scalability}
	\end{centering}
\end{figure}

\subsection{Calibration of Liquid Drop Model}\label{subsec:application_LDM}
Over the past decade or so, the statistical tools of uncertainty quantification have experienced a robust rump-up in use in the field of nuclear physics \citep{Ireland2015}. Bayesian calibration has been especially popular because it enhances the understanding of nuclear model's structure through parameter estimation and potentially advances the quality of nuclear modeling by accounting for systematic errors. In this context, we use our variational Algorithm \ref{alg:TruncatedDfinal} to calibrate the 4-parameter Liquid Drop Model (LDM)  \citep{Myers1966,Kirson2008,Benzaid2020} which is a global (across the whole nuclear chart) model of nuclear binding energies; the minimum energy needed to disassemble the nucleus of an atom into free protons and neutrons. Nuclear binding energyis equivalent (energy-mass equivalence explained by $E=mc^2$) to the mass defect that corresponds to the difference between the mass number of a nucleus and its actual measured mass. In principle, the LDM treats the nucleus like molecules in a drop of incompressible fluid of very high density. Despite this simplification, the LDM accounts for the spherical shape of most nuclei and makes reasonable estimates of average properties of nuclei. The LDM is formulated through the semi-empirical mass formula as:
\begin{equation}\label{eqn:LDM}
\begin{split}
&E_{\rm B}(N,Z) = \theta_{\rm vol}A - \theta_{\rm surf}A^{2/3} - \theta_{\rm sym} \frac{(N-Z)^2}{A} - \theta_{\rm C} \frac{Z(Z-1)}{A^{1/3}}.
\end{split}
\end{equation}
where $Z$ is the proton number, $N$ is the neutron number, and  $A=Z+N$ is the mass number of an atom. The calibration parameters are $\thetab = (\theta_{\rm vol}, \theta_{\rm surf}, \theta_{\rm sym}, \theta_{\rm C})$ representing the volume, surface, symmetry and Coulomb energy, respectively. These parameters have specific physical meaning, where $\theta_{\rm vol}$ is proportional to the volume of the nucleus for instance. See \cite{krane1987introductory} for more details. Here we note that this is by no means the first case when Bayesian methodology is applied to study the LDM. In fact, the LDM is a popular model for statistical application \citep{Bertsch2005, Yuan2016, Bertsch17, kejzlar2021} which is why we choose the model to illustrate our methodology as well. The LDM also generally performs better on heavy nuclei as compared to the light nuclei which alludes to the existence of a significant systematic discrepancy between the model and the experimental binding energies \citep{Reinhard2006, kejzlar2020statistical}. Namely, we consider the following statistical model
\begin{equation}\label{eqn:VBI:LDMfull}
y = E_{\text{B}}(N,Z) + \delta(N,Z) + \sigma \epsilon,
\end{equation}
where $\delta(N,Z)$ represents the unknown systematic discrepancy between the semi-empirical mass formula and the experimental binding energies $y$. The parameter $\sigma$ is as usual the scale of observation error $\epsilon \sim \cN(0,1)$. The nuclear physics community often \citep{Dobaczewski2014} considers the least squares (LS) estimator of $\thetab$ defined as 
\begin{equation}\label{eqn:chisq}
\hat{\thetab}_{L_2}= \argmin_{\thetab}
\sum_{i=1}^n\left(y_i - E_{\rm B}(N_i,Z_i)  \right)^2
\end{equation}
which is also the maximum likelihood estimate of $\thetab$ in the case of $\delta = 0$. The benefit of this estimator is that it is fast, easy-to-compute, and allows for analysis under the standard linear regression theory. It, however, neglects some sources of uncertainty that are accounted for in the Bayesian calibration framework.

\begin{figure}[h] 
	\begin{centering}
		\includegraphics[width=0.8\textwidth]{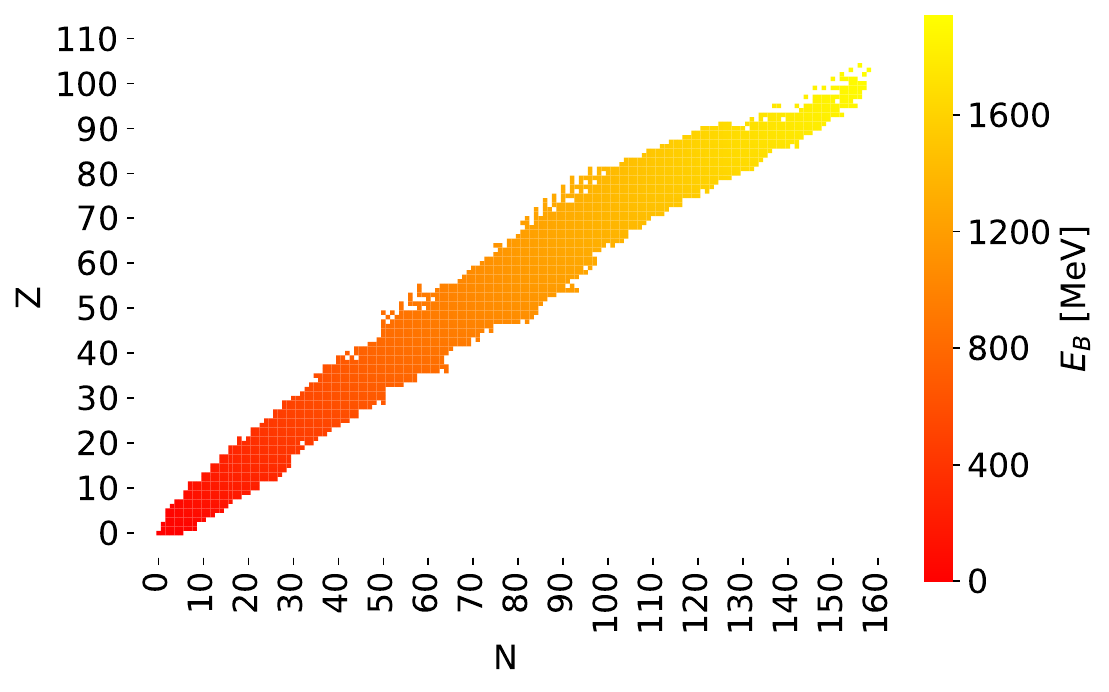}
		\caption{Experimental binding energies of nuclei in AME2003 dataset (2225 observations).}
		\label{fig:VBI:AME2003}
	\end{centering}
\end{figure}

To this end, we shall consider a GP prior with the mean zero and the squared exponential covariance function for the systematic discrepancy $\delta(Z,N)$. Since the main purpose of the example is to provide a canonical illustration of the methodology in a real data scenario, we also set a GP prior for the LDM and treat $E_B(Z,N)$ as an unknown function. We use 2000 experimental binding energies randomly selected from the AME2003 dataset \citep{AME2003} (publicly available at \url{http://amdc.impcas.ac.cn/web/masstab.html}) for calibration, see Figure \ref{fig:VBI:AME2003}, and an additional set of $10^4$ model evaluations.
The calibration inputs were generated with the Latin hypercube design so that all the reasonable values of $(\theta_{\rm vol}, \theta_{\rm surf}, \theta_{\rm sym}, \theta_{\rm C})$ given by the literature are covered \citep{Weizsacker1935,Bethe36,Myers1966,Kirson2008,Benzaid2020}. The model inputs $(Z,N)$ were selected from the set of 2000 experimental binding energies, duplicated five-fold, and randomly permutated among the generated calibration inputs to span only the set of relevant nuclei. This relatively large number of model runs was chosen so that the combined 6 dimensional space of calibration parameters and model inputs is sufficiently covered considering the existence of a non-trivial systematic discrepancy. In fact, the uniform experimental design would amount only to 4-5 points per dimension.

Independent Gaussian distributions centered at the LS estimates $\hat{\thetab}_{L_2}$ (in Table \ref{tab:LDM_results}) with standard deviations large enough to cover the space of inputs used for generating the model runs were selected to represent the prior knowledge about the calibration parameters. Independent gamma distributions were used as the prior models for the hyperparameters of the GP's covariance functions. We choose the variational family to be fully-factorized with the Gaussian distributions for real valued parameters and the gamma distributions for positive variables. The means of variational families were initialized as random samples from their respective prior distributions and the variances were set to match those of the prior distributions. We used the AdaGrad for stochastic optimization. See Appendix \ref{app:LDM} for further discussion on the prior distributions and experimental design.

\subsubsection{Results} \label{subsubsec:application_LDM_results}
Including the generated model runs, the overall size of training dataset is $1.2 \times 10^4$ which already makes the MCMC based Bayesian calibration impractical, as illustrated by the simulation study in Section \ref{subsec:application_simulation}. We therefore asses the quality of variational approximation only against the LDM with the standard LS estimation and do not consider the MCMC methods. In particular, we consider the testing dataset of the remaining $225$ experimental binding energies in AME2003 that were excluded from the training data. The predictions $\hat{\yb}^*$ of these testing binding energies $\yb^*$ were calculated, under the variational approximation, as the posterior means of $\yb^*$ conditioned on the $1.2 \times 10^4$ binding energies from the training data set, i.e., the posterior means of the predictive distribution $p(\yb^*|\db)$. The predictions under the LS estimates $\hat{\thetab}_{L_2}$ were given by the  semi-empirical mass formula  \eqref{eqn:LDM}.

\begin{figure*}[h!] 
	\begin{centering}
		\includegraphics[width=0.9\textwidth]{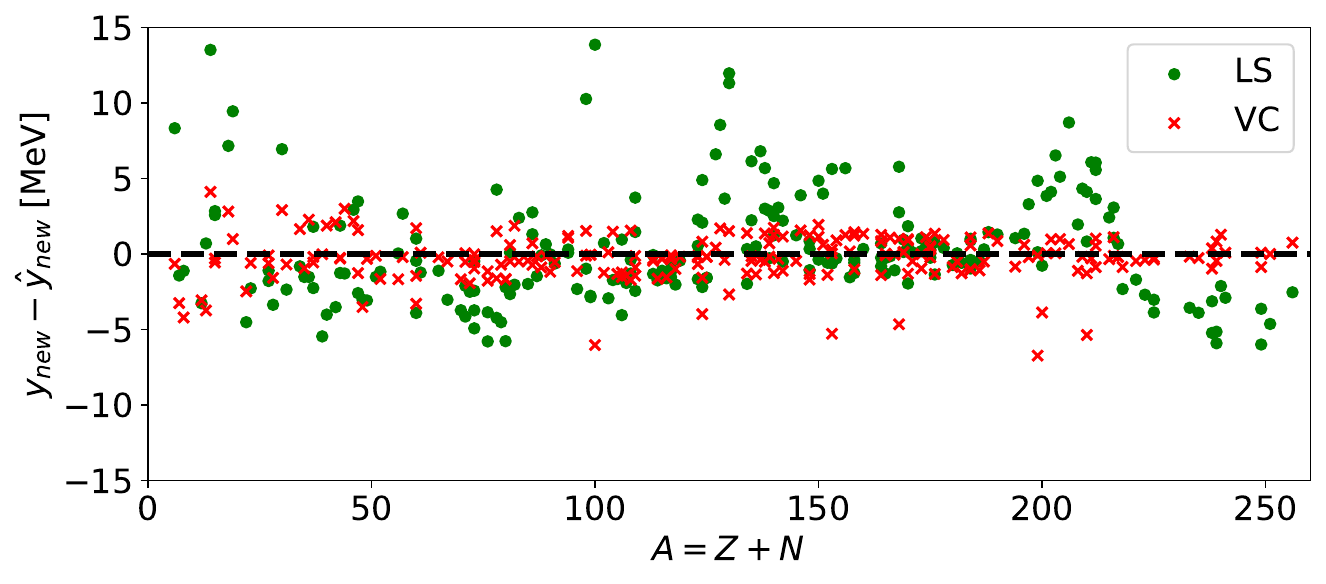}
		\caption{The residual plot for 225 experimental binding energies in the testing dataset.}
		\label{fig:VBI:LDM:deviation}
	\end{centering}
\end{figure*}

Table \ref{tab:LDM_results} gives the root MSE for both methods under consideration. The VC (Algorithm \ref{alg:TruncatedDfinal}) results are based on a 24 hour window dedicated to running the algorithm with 50 samples used to approximate the expectations, 10 samples used to implement the control variates, and the truncation level selected to be $l = 3$. By using GPs to account for the systematic discrepancies of the semi-empirical mass formula and the uncertainty of the LDM itself, we were able to significantly reduce the root MSE approx. $57\%$ compared to the LS benchmark. Table \ref{tab:LDM_results} additionally shows the calibration parameter estimates and their standard errors. The estimates under the VC are given by the means of their variational families. Both the methods calibrate the LDM around the same values with notably low standard errors of the LS estimates. This is, however, expected since $\hat{\thetab}_{L_2}$ are ordinary LS estimates that in the presence of heteroscedasticity (see Figure \ref{fig:VBI:LDM:deviation}) become inefficient and tend to significantly underestimate the true variance \citep{goldberger1966econometric, johnston1976econometric}.
\begin{table*}[h]
	\begin{tabular}{l|llll|c}
		\hline 
		\hline 
		Method  & \multicolumn{4}{|c|}{Parameter estimate and standard errors} & Testing error \\ \cline{2-6}
		& $a_{vol}$ & $a_{surf}$ & $a_{sym}$ & $a_C$ & $\sqrt{MSE}$ (MeV)\\ \hline
		LS & 15.42 (0.027) & 16.91 (0.086) & 22.47 (0.070) & 0.69 (0.002) & 3.54\\
		VC & 15.78 (0.198) & 15.99 (0.681) & 21.94 (0.510) & 0.68 (0.018) & 1.52  \\ \hline\hline 
	\end{tabular}
	\caption{The root MSE of the VC (Algorithm \ref{alg:TruncatedDfinal}) after 24 hours dedicated to running the algorithm compared with the root MSE based on the LS estimates. The parameter estimates (and their standard errors) are also displayed. \label{tab:LDM_results}}
\end{table*}

The residual plot in Figure \ref{fig:VBI:LDM:deviation}, showing the difference between $\yb^*$ and $\hat{\yb}^*$ as a function of the nuclear mass number $A$, clearly demonstrates a better fit of the testing data with our methodology than is achieved by the simple LS fit. The majority of the residuals appear to be randomly spread around 0 which strongly supports the efficiency of GPs in accounting for the systematic discrepancy between the model and the physical process.

\section{Discussion}\label{sec:conclusion}
We developed and studied a VBI based approach to Bayesian calibration of computer models under the celebrated framework of \cite{KoH} which has been heavily utilized by practitioners for almost two decades. Our method consists of scalable and statistically principled tools for UQ of computationally complex and many-parameter computer models. We exploit the probabilistic theory of approximation coupled with pairwise construction of multivariate copulas using truncated regular vines to establish these tools. The theoretical justification for scalability was also discussed. Here we also note that while our work mainly focused on the model calibration framework, we can use the VBI algorithm developed in this paper for any models with complex Gaussian likelihoods with minimal modifications needed. Additionally, we dedicated a significant portion of this text to the description of implementation details that are often neglected in the literature. We discussed the choice of learning rate for stochastic optimization and outlined techniques to reduce the variance of noisy gradient estimates which include the Rao-Blackwellization, control variates, and importance sampling.

We are aware of efficient modeling of covariance modeling in spatial statistics that facilitates interpolation and prediction in the spatial domain. Our objective here is not to model spatial covariance function instead build variational Bayes inference for calibrated computer models. The issues in calibrated computer models are fundamentally different. Therefore, the standard practice of spatial statistics is not directly importable for inference. We anticipate this research will have an impact on Bayesian spatial statistics in the near future.

In our examples, we first carried out an extensive simulation study that provided empirical evidence for accuracy and scalability of our method in scenarios where traditional MCMC based approaches become impractical. We established the superiority of the VC over the MH algorithm and the NUTS in terms of time efficiency and memory requirements. We also demonstrated the opportunities given by our method for practitioners on a real data example through calibration of the Liquid Drop Model of nuclear binding energies.

There are a few natural directions to enhance the methodology provided in this work from both computational and theoretical perspectives. First, an {\it a priori} method to select a sufficient truncation level for vine copulas would be beneficial to avoid the current sequential approach. For example, \cite{BRECHMANN201519} discuss the use of fit indices for finding sufficient truncation. Secondly, the theoretical justification for our method would greatly benefit from establishing the link between the ELBO and the l-truncated ELBO which is the ultimate driving force behind the computational efficiency of the VC. Additionally, there are other alternatives to the traditional MCMC than VBI that have shown to be effective in handling massive datasets. The stochastic gradient MCMC \citep{SGMCMC2015} algorithm, for instance, utilizes similar data subsampling trick as VBI (see Section \ref{sec:vbi} for details) which has been successfully applied in deep learning \citep{Feming2019} or state space models \citep{SGMCMCSP2019}. A similar copula likelihood decomposition to the one proposed in this paper could be used for computer model calibration via stochstic gradient MCMC, however, it would require a non-trivial algorithmic development that is beyond the scope of this work.

\begin{appendices}
	\section{Scalable Algorithm with Truncate C-Vine Copulas}\label{app:C_vine}
Here we present the details of the C-vine based versions of Algorithm~\ref{alg-truncated} and Algorithm~\ref{alg:TruncatedDfinal}. First, we can decompose the log-likelihood $\log p(\bm{d}|\phib)$ using a C-vine as
\begin{equation}\label{egn:VBI:pCvine}
	\log p(\bm{d}|\phib) = \sum_{j = 1}^{N-1} \sum_{i = 1}^{N-j} p^C_{j, j+i}(\phib),   
\end{equation}
where
\begin{equation}\label{eqn:VBI:PcDefinition}
\begin{split}
    p^C_{j, j +i}(\phib) &=  \log c_{j,(j+i); 1, \dots, (j-1)} + \frac{1}{N-1}\big(\log p_j(d_j|\phib) + \log p_{j+i}(d_{j+i}|\phib)\big).
\end{split}
\end{equation}
This now yields the following expression for the ELBO gradient:
\begin{equation}\label{eqn:VBI:ELBOCvine}
\begin{split}
	\nabla_\lambda \mathcal{L}(\lambda) &= \sum_{j = 1}^{N-1} \sum_{i = 1}^{N-j}\mathbb{E}_q\bigg[\nabla_{\lambdab} \log q(\phib|\lambdab)(	p^C_{j, j+i}(\phib))\bigg] - \mathbb{E}_q\bigg[\nabla_{\lambdab} \log q(\phib|\lambdab)\log\frac{ q(\phib|\lambdab)}{p(\phib)}\bigg].
	\end{split}
\end{equation}
Equivalently to Proposition \ref{propostion:Dvine}, we have the following proposition that establishes the noisy unbiased estimate of the gradient \eqref{eqn:VBI:ELBOCvine} using the C-vine copula decomposition.
\begin{proposition}\label{propostion:Cvine}
\textit{Let $\tilde{\mathcal{L}}_C(\lambdab)$ be an estimate of the ELBO gradient $\nabla_{\lambdab} \mathcal{L}(\lambdab)$ defined as
\begin{equation*}\label{eqn:VBI:ELBOCvineNoisy}
\begin{split}
\tilde{\mathcal{L}}_C(\lambdab) &= \frac{N(N-1)}{2}\mathbb{E}_q\bigg[\nabla_{\lambdab}\log q(\phib|\lambdab)(	p^C_{I_C(K)}(\phib))\bigg] - \mathbb{E}_q\bigg[\nabla_{\lambdab} \log q(\phib|\lambdab)\log \frac{ q(\phib|\lambdab)}{ p(\phib)}\bigg],
\end{split}
\end{equation*}
where $K \sim U(1, \dots, \frac{N(N-1)}{2})$, and $I_C$ is the bijection
\begin{equation*}\label{eqn:VBI:bijectionCvine}
\begin{split}
&I_C:\{1, \dots, \frac{N(N-1)}{2}\} \rightarrow \{(j,j+i): i \in \{1, \dots, N-j\} \text{ for } j \in \{1, \dots N-1\}\},
\end{split}
\end{equation*}
then $\tilde{\mathcal{L}}_C(\lambdab)$ is unbiased i.e., $\mathbb{E}(\tilde{\mathcal{L}}_C(\lambdab)) = \nabla_{\lambdab} \mathcal{L}(\lambdab)$.}
\end{proposition}

Again, $\tilde{\mathcal{L}}_C(\lambdab)$ can be relatively costly to compute for large datasets due to the recursive nature of the copula density computations. We now carry out exactly the same development an using l-truncated C-vine as in the case of Proposition \ref{proposition:DvineTrunc} and Proposition \ref{propostion:DvineTruncUnbiase}.

\begin{proposition}\label{proposition:CvineTrunc}
\textit{If the copula of $p(\bm{d}|\phib)$ is distributed according to an l-truncated C-vine, we can rewrite}
\begin{equation}\label{eqn:CvineTruncatedP}
\log p(\bm{d}|\phib) = \sum_{j = 1}^{l} \sum_{i = 1}^{N-j} p^{C_l}_{i, i+j}(\phib),
\end{equation}
\textit{where}
\begin{equation}\label{eqn:VBI:pDefCvine}
\begin{split}
    p^{c_l}_{i, i+j}(\phib) &=   \log c_{j,(j+i); 1, \dots, (j-1)}+ \frac{1}{a_j}\log p_j(d_j|\phib) + \frac{1}{b_{j+i}} \log p_{j+i}(d_{j+i}|\phib),
    \end{split}
\end{equation}
\textit{and}
\begin{align*}\label{eqn:VBI:aibiCvine}
a_j &=N-1, \\
b_{j+i} &= (N-1-l)\mathbbm{1}_{j + i \le l} + l.
\end{align*}
\end{proposition}
Let us now replace the full log-likelihood $\log(\bm{d}|\phib)$ in the definition of ELBO with the likelihood based on a truncated vine copula. This yields the l-truncated ELBO for the l-truncated C-vine
\begin{equation}\label{eqn:VBI:ELBOCvineTruncated}
\mathcal{L}_{C_l}(\lambdab) = \mathbb{E}_q\bigg[\sum_{j = 1}^{l} \sum_{i = 1}^{N-j} p^{C_l}_{j, j+i}(\phib)\bigg] - KL(q(\phib|\lambdab)||p(\phib))
\end{equation}
with its gradient
\begin{equation*}\label{eqn:VBI:ELBOCvineLgrad}
\begin{split}
\nabla_{\lambdab} \mathcal{L}_{C_l}(\lambdab) &= \sum_{j = 1}^{l} \sum_{i = 1}^{N-j}\mathbb{E}_q\bigg[\nabla_{\lambdab} \log q(\phib|\lambdab)(p^{C_l}_{j, j+i}(\phib))\bigg] -\mathbb{E}_q\bigg[\nabla_{\lambdab} \log q(\phib|\lambdab)\log\frac{q(\phib|\lambdab)}{p(\phib)}\bigg].
\end{split}
\end{equation*}
Consequently, we get the following proposition that establishes the noisy unbiased estimate of $\nabla_{\lambdab} \mathcal{L}_{C_l}(\lambdab)$.
\begin{proposition}\label{propostion:CvineTruncUnbiase}
\textit{Let $\tilde{\mathcal{L}}_{C_l}(\lambdab)$ be an estimate of the ELBO gradient $\nabla_{\lambdab} \mathcal{L}_{C_l}(\lambdab)$ defined as
\begin{equation*}\label{eqn:VBI:ELBOCvineTruncNoisy}
\begin{split}
&\tilde{\mathcal{L}}_{C_l}(\lambdab) = \frac{l(2N-(l + 1))}{2}\mathbb{E}_q\bigg[\nabla_{\lambdab} \log q(\phib|\lambdab)(	p^{C_l}_{I_{C_l}(K)}(\phib))\bigg] - \mathbb{E}_q\bigg[\nabla_{\lambdab} \log q(\phi|\lambdab)\log\frac{ q(\phi|\lambdab)}{ p(\phi)}\bigg],
\end{split}
\end{equation*}
where $K \sim U(1, \dots, \frac{l(2N-(l + 1))}{2})$, and $I_{C_l}$ is the bijection
\begin{equation*}\label{eqn:VBI:bijectionCvineTrunc}
\begin{split}
&I_{C_l}:\{1, \dots, \frac{l(2N-(l + 1))}{2}\} \rightarrow \{(j,j+i): i \in \{1, \dots, N-j\} \text{ for } j \in \{1, \dots l\}\},
\end{split}
\end{equation*}
then $\tilde{\mathcal{L}}_{C_l}(\lambdab)$ is unbiased i.e., $\mathbb{E}(\tilde{\mathcal{L}}_{C_l}(\lambdab)) = \nabla_{\lambdab} \mathcal{L}_{C_l}(\lambdab)$.}
\end{proposition}
Algorithm \ref{alg:TruncatedCbasic} postulates the version of Algorithm \ref{alg-truncated} based on the truncated C-vine decomposition.
	
	\begin{algorithm}[h]
\DontPrintSemicolon
	\caption{Variational calibration with truncated C-vine copulas.\label{alg:TruncatedCbasic}}
	\addtocontents{loa}{\vskip 12pt}
		\KwIn{Data $\bm{d}$, mean and covariance functions for GPs, variational family $q(\phib|\lambdab)$, \textbf{truncation level l}}
		$\lambda \leftarrow$ random initial value\;
		$t \leftarrow 1$\;
		\Repeat{change of $\lambdab$ is less than $\epsilon$}{
		\For{$s = 1$ to $S$}{
		$\phib[s] \sim q(\phib|\lambdab)$
		}
		$K \leftarrow U(1, \dots, \frac{l(2N-(l + 1))}{2})$\;
		$\rho \leftarrow$ $t^{\textrm{th}}$ value of a Robbins-Monro sequence\;
		$
		\lambdab \leftarrow \lambdab + \rho \frac{1}{S}\sum_{s=1}^{S}\bigg[\frac{l(2N-(l + 1))}{2}\nabla_{\lambdab} \log q(\phib[s]|\lambdab) \times \big(	p^{C_l}_{I_{C_l}(K)}(\phib[s]) -\frac{2}{l(2N-(l + 1))} \log\frac{ q(\phib[s]|\lambdab)}{p(\phib[s])}\big)\bigg]
		$\;
		$t \leftarrow t + 1$\;}
\end{algorithm}

\subsection{Variance Reduction}\label{app:sub_C_vine}
	Let us now consider the MC approximation of the gradient estimator $\tilde{\mathcal{L}}_{C_l}(\lambdab)$. The $j^{th}$ entry of the estimator with Rao-Blackwellization is
\begin{equation*}
\begin{split}
&\frac{1}{S}\sum_{s=1}^{S}\bigg[\frac{l(2N-(l + 1))}{2}\nabla_{\lambdab_j} \log q(\phi_j[s]|\lambdab_j)\big(\tilde{p}_{(j)}(\phi[s]) - \frac{2}{l(2N-(l + 1))}\log\frac{ q(\phi_j[s]|\lambdab_j)}{p(\phi_j[s])}\big)\bigg],
\end{split}
\end{equation*}
where $\tilde{p}_{(j)}(\phib)$ are here the components of $p^{C_l}_{I_{C_l}(K)}(\phib)$ that include $\phi_j$.

We can again use the control variates to reduce the variance of MC approximation of the gradient estimator $\tilde{\mathcal{L}}_{C_l}(\lambdab)$. In particular, we consider the following $j^{th}$ entry of the Rao-Blackwellized MC approximation of the gradient estimator $\tilde{\mathcal{L}}_{C_l}(\lambdab)$ with control variates
	\begin{equation*}\label{eqn:CVELBOCvine}
	\begin{split}
	&\tilde{\mathcal{L}}_{C_l}^{CV(j)}(\lambdab) = \sum_{s=1}^{S}\bigg[\frac{l(2N-(l + 1))}{2S}\nabla_{\lambdab_j} \log q(\phi_j[s]|\lambdab_j)(	\tilde{p}_{(j)}(\phi[s]) - \frac{2(\log\frac{ q(\phi_j[s]|\lambdab_j)}{p(\phi_j[s])} + \hat{a}^C_j)}{l(2N-(l + 1))})\bigg],
	\end{split}
\end{equation*}
where $\hat{a}^C_j$ is the estimate of the optimal control variate scalar $a^*$ based on $S$ (or fever) independent draws from the variational distribution. Namely,
	\begin{equation*}
	\hat{a}^C_j = \frac{\widehat{\mathbb{C}ov}_q(\xi^C(\phib),\psi^C(\phib))}{\widehat{\mathbb{V}ar}_q(\psi^C(\phib))},
\end{equation*}
where
\begin{align*}
\begin{split}
  \xi^C(\phib) &=   \frac{l(2N-(l + 1))}{2}\nabla_{\lambdab_j} \log q(\phi_j|\lambdab_j)\bigg(\tilde{p}_{(j)}(\phib) -\frac{2 \log q(\phi_j|\lambdab_j)}{l(2N-(l + 1)) \log p(\phi_j)}\bigg)
  \end{split}
\end{align*}
and $\psi^C(\phib) =   \nabla_{\lambdab_j} \log q(\phi_j|\lambdab_j)$.
	
As in the case of the D-vine, we now derive the ultimate  Algorithm~\ref{alg:TruncatedCfinal}. Again, instead of taking the samples from $q(\phib| \lambdab)$ to approximate the gradient estimates, we will take samples from an overdispersed distribution $r(\phib|\lambdab, \tau)$. Combining the Rao-Blackwellization, control variates, and importance sampling, we have the following $j^{th}$ entry of the MC approximation of the gradient estimator $\tilde{\mathcal{L}}_{C_l}(\lambda)$ 
\begin{align*}\label{eqn:CVELBOCimp}
&\tilde{\mathcal{L}}_{C_l}^{OCV(j)}(\lambdab)=\sum_{s=1}^{S}\bigg[\frac{l(2N-(l + 1))}{2S}\nabla_{\lambdab_j} \log q(\phi_j[s]|\lambdab_j)(\tilde{p}_{(j)}(\phi[s])-\frac{2(\log\frac{ q(\phi_j[s]|\lambdab_j)}{p(\phi_j[s])} + \tilde{a}^C_j)}{l(2N-(l + 1))})w(\phi_j[s])\bigg],
\end{align*}
where $\phib[s] \sim r(\phib|\lambdab, \tau)$ and $w(\phib[s] ) = q(\phib[s] |\lambdab) / r(\phib[s]|\lambdab, \tau)$ with
\begin{equation*}
	\tilde{a}^C_j = \frac{\widehat{\mathbb{C}ov}_q(\tilde{\xi}^C(\phib),\tilde{\psi}^C(\phib))}{\widehat{\mathbb{V}ar}_q(\tilde{\psi}^C(\phib))},
\end{equation*}
where
\begin{align*}
\begin{split}
  \tilde{\xi}^C(\phib) &=   \frac{l(2N-(l + 1))w(\phi_j)}{2}\nabla_{\lambdab_j} \log q(\phi_j|\lambdab_j)\bigg(	\tilde{p}_{(j)}(\phib) -  \frac{2\log\frac{ q(\phi_j|\lambdab_j)}{p(\phi_j)}}{l(2N-(l + 1))}\bigg)
  \end{split}
\end{align*}
and $\tilde{\psi}^C(\phib) =   \nabla_{\lambdab_j} \log q(\phi_j|\lambdab_j) w(\phi_j)$.

\begin{algorithm}[h]
\DontPrintSemicolon
	\caption{Variational calibration with truncated C-vine copulas II.\label{alg:TruncatedCfinal}}
	\addtocontents{loa}{\vskip 2pt}
		\KwIn{Data $\bm{d}$, mean and covariance functions for GPs, variational family $q(\phib|\lambdab)$, dispersion parameter $\tau$, and \textbf{truncation level l}}
		$\lambda \leftarrow$ random initial value\;
		$t \leftarrow 1$\;
		\Repeat{change of $\lambdab$ is less than $\epsilon$}{
		\For{$s = 1$ to $S$}{
		$\phib[s] \sim r(\phib|\lambdab, \tau)$\tcp*{Random sample from $r$}
		}
		$K \leftarrow U(1, \dots, \frac{l(2N-(l + 1))}{2})$\;
		$\rho \leftarrow$ $t^{\textrm{th}}$ value of a Robbins-Monro sequence\;
		$
		\lambdab \leftarrow \lambdab + \bm{\rho}  \sum_{s=1}^{S}\bigg[\frac{l(2N-(l + 1))}{2S}\nabla_{\lambdab_j} \log q(\phi_j[s]|\lambdab_j) \times \big(\tilde{p}_{(j)}(\phib[s])-\frac{2(\log\frac{ q(\phi_j[s]|\lambdab_j)}{p(\phi_j[s])} + \tilde{a}^C_j)}{l(2N-(l + 1))}\big)w(\phi_j[s])\bigg]
		$\;
		$t \leftarrow t + 1$\;}
\end{algorithm}

\section{Proofs}\label{sec:VBI:details:proofs}
\paragraph{Proof of Proposition \ref{propostion:Dvine}.}
Since $P(K = k) = \frac{2}{N(N-1)}$, we have directly from the definition of expectation
\begin{align*}
    &\mathbb{E}(\tilde{\mathcal{L}}_D(\lambdab)) =\frac{N(N-1)}{2} \sum_{k=1}^{\frac{N(N-1)}{2}}\frac{2}{N(N-1)}\mathbb{E}_q\bigg[\nabla_{\lambdab}\log q(\phib|\lambdab)(	p^D_{I_D(k)}(\phib))\bigg] \\
    &- \mathbb{E}_q\bigg[\nabla_{\lambdab} \log q(\phib|\lambdab)\log \frac{ q(\phib|\lambdab)}{ p(\phib)}\bigg] = \nabla_{\lambdab} \mathcal{L}(\lambdab).
\end{align*}
The final equality is the consequence of the uniqueness of the pairs of variables in the conditioned sets of the copula density $c_{i,(i+j); (i+1), \dots, (i+j-1)}$, and that $\frac{N(N-1)}{2}$ is the number of unordered pairs of $N$ variables.

\paragraph{Proof of Proposition \ref{proposition:DvineTrunc}.}
It is sufficient to show that for $l \in \{1, \dots, N-1\}$ the following equality holds:
\begin{equation}\label{eqn:proposition:D}
\begin{split}
    &\sum_{j = 1}^{l} \sum_{i = 1}^{N-j}\bigg[\frac{1}{a_i}\log p_i(d_i|\phib) + \frac{1}{b_{i+j}} \log p_{i+j}(d_{i+j}|\phib)\bigg] = \sum_{k = 1}^N \log p(\db_k|\phib),
    \end{split}
\end{equation}
where
\begin{align*}
a_i &= 2l - \bigg[(l+1-i)\mathbbm{1}_{i \le l} + (l - N +i)\mathbbm{1}_{i > N - l}\bigg], \\
b_{i+j} &= 2l - \bigg[(l+1-j-i)\mathbbm{1}_{i + j \le l} +  (l - N +j+i)\mathbbm{1}_{i + j > N - l}\bigg].
\end{align*}
To show this, let us consider the summation
\begin{align*}
    &\sum_{j = 1}^{l} \sum_{i = 1}^{N-j}\bigg[\log p_i(d_i|\phib) +  \log p_{i+j}(d_{i+j}|\phib)\bigg]\\
    &=\sum_{j = 1}^{l} \bigg[(\log p_1(d_1|\phib) +  \log p_{1+j}(d_{1+j}|\phib)) + \dots  +(\log p_{N-j}(d_{N-j}|\phib) +  \log p_{N}(d_{N}|\phib))\bigg].
\end{align*}
For $l = 1$, we get
\begin{align*}
    &\sum_{j = 1}^{l} \sum_{i = 1}^{N-j}\bigg[\log p_i(d_i|\phib) +  \log p_{i+j}(d_{i+j}|\phib)\bigg]\\
    &=(\log p_1(d_1|\phib) +  \log p_{2}(d_{2}|\phib)) + \dots  +(\log p_{N-1}(d_{N-1}|\phib) +  \log p_{N}(d_{N}|\phib)),
\end{align*}
and for $l \ge 2$
\begin{align*}
    &\sum_{j = 1}^{l} \sum_{i = 1}^{N-j}\bigg[\log p_i(d_i|\phib) +  \log p_{i+j}(d_{i+j}|\phib)\bigg]\\
    &=\bigg[(\log p_1(d_1|\phib) +  \log p_{2}(d_{2}|\phib)) + \dots +(\log p_{N-1}(d_{N-1}|\phib) +  \log p_{N}(d_{N}|\phib))\bigg] +\dots \\ &\qquad + \bigg[(\log p_1(d_1|\phib) +  \log p_{1+l}(d_{1+l}|\phib)) + \dots +(\log p_{N-l}(d_{N-l}|\phib) +  \log p_{N}(d_{N}|\phib))\bigg].
\end{align*}
Note that in the case of $l = N-1$, the last summation consists of only one element $\log p_1(d_1|\phib) +  \log p_{1+l}(d_{1+l}|\phib)$. By careful examination of the two cases above, we get the following results. For $2l \le N$:
\begin{align*}
    &\sum_{j = 1}^{l} \sum_{i = 1}^{N-j}\bigg[\log p_i(d_i|\phib) +  \log p_{i+j}(d_{i+j}|\phib)\bigg]\\
    &=\sum_{k = 1}^l (l + k -1)\log p_k(d_k|\phib) + \sum_{k =l + 1}^{N-l} 2l \log p_k(d_k|\phib) +\sum_{k =N - l + 1}^{N} (N - i + l)\log p_k(d_k|\phib),
\end{align*}
where the middle term disappears in the case $2l = N$, and for $2l > N$:
\begin{align*}
    &\sum_{j = 1}^{l} \sum_{i = 1}^{N-j}\bigg[\log p_i(d_i|\phib) +  \log p_{i+j}(d_{i+j}|\phib)\bigg]\\
    &=\sum_{k = 1}^{N-l} (l + k -1)\log p_k(d_k|\phib) +\sum_{k = N - l + 1}^{l} (N-1) \log p_k(d_k|\phib) + \sum_{k =l+1}^{N} (N - i + l)\log p_k(d_k|\phib).
\end{align*}
If we now check that $a_i$ equals to the factors in front of the log-likelihoods in the two cases above, the proof of Proposition \ref{proposition:DvineTrunc} is complete. Note that once we check the equality for $a_i$, the same directly translates to $b_{i+j}$ since $b_{i+j}$ is $a_i$ with indices set to $i+j$ instead of $i$. Indeed, for $2l \le N$
\begin{equation*}
a_{i} = 
   \begin{cases} 
      l +i -1 & i\le l\\
      2l &  l < i \le N -l\\
      N - i + l &  N - l < i
   \end{cases},
\end{equation*}
and for $2l > N$
\begin{equation*}
a_{i} = 
   \begin{cases} 
      l +i -1 & i\le N - l\\
      N -1 &  N - l < i \le l\\
      N - i + l &  l < i
   \end{cases}.
\end{equation*}

\paragraph{Proof of Proposition \ref{propostion:DvineTruncUnbiase}.}
By the construction of R-vine (see \cite{Kurowicka2006}), each tree $\mathcal{T}_i$, for $i = 1, \dots, N-1$ has exactly $N-i$ edges (these are the unique conditioned variable pairs). For any R-vine truncated at level $l \in \{1, \dots, N-1\}$, we get the number of edges to be
\begin{equation*}
\sum_{i = 1}^l (N - i) = lN - \frac{l(l+1)}{2} = \frac{l(2N - (l+1))}{2}
\end{equation*}
The rest of the proof is identical with that of Proposition \ref{propostion:Dvine} due to the uniqueness of the conditioned variable pairs in the copula density $c_{i,(i+j); (i+1), \dots, (i+j-1)}$, but in this case $P(K = k) = \frac{2}{l(2N - (l+1))}$.

\paragraph{Proof of Proposition \ref{propostion:Cvine}.}
The proof is identical with that of Proposition \ref{propostion:Dvine} since each conditioned pair in the copula density $c_{j,(j+i); 1, \dots, (j-1)}$ is unique as well.

\paragraph{Proof of Proposition \ref{proposition:CvineTrunc}.}
It is sufficient to show that for $l \in \{1, \dots, N-1\}$ the following equality holds:
\begin{equation}\label{eqn:proposition:C}
\begin{split}
    &\sum_{j = 1}^{l} \sum_{i = 1}^{N-j}\bigg[\frac{1}{a_j}\log p_j(d_j|\phib) + \frac{1}{b_{j+i}} \log p_{j+i}(d_{j+i}|\phib)\bigg] = \sum_{k = 1}^N \log p(\db_k|\phib),
    \end{split}
\end{equation}
where
\begin{align*}
a_j &= N-1, \\
b_{j + i} &= (N-1-l)\mathbbm{1}_{j + i \le l} + l.
\end{align*}
To show this, let us consider the following summation
\begin{align*}
        &\sum_{j = 1}^{l} \sum_{i = 1}^{N-j}\bigg[\log p_j(d_j|\phib) + \log p_{j+i}(d_{j+i}|\phib)\bigg]=\\ 
        & \sum_{j = 1}^{l} \bigg[(N - j)\log p_j(d_j|\phib) + \sum_{i = 1}^{N-j}\log p_{j+i}(d_{j+i}|\phib)\bigg] =\\
        &  \sum_{j = 1}^{l}(N - j)\log p_j(d_j|\phib) + \sum_{j = 1}^{l}\bigg[\log p_{j+1}(d_{j+i}|\phib)) + \dots + \log p_{N}(d_{N}|\phib)\bigg].
\end{align*}
Now, for $l = 1$, we have
\begin{align*}
    &\sum_{j = 1}^{l}\bigg[\log p_{j+1}(d_{j+1}|\phib)) + \dots + \log p_{N}(d_{N}|\phib)\bigg]
    = \log p_{2}(d_{2}|\phib) + \dots \log p_{N}(d_{N}|\phib).
\end{align*}
For $l \ge 2$, we have
\begin{align*}
    &\sum_{j = 1}^{l}\bigg[\log p_{j+1}(d_{j+1}|\phib)) + \dots + \log p_{N}(d_{N}|\phib)\bigg]=\\
    & \bigg[\log p_{2}(d_{2}|\phib) + \dots \log p_{N}(d_{N}|\phib)\bigg] + \dots + \bigg[\log p_{l+1}(d_{l+1}|\phib) + \dots \log p_{N}(d_{N}|\phib)\bigg] .
\end{align*}

Therefore we can rewrite
\begin{align*}
 &\sum_{j = 1}^{l}\bigg[\log p_{j+1}(d_{j+1}|\phib)) + \dots + \log p_{N}(d_{N}|\phib)\bigg]= \sum_{j = 1}^{l}(j-1) \log p_j(d_j|\phib) + \sum_{j = l+1}^{N}l \log p_j(d_j|\phib) 
\end{align*}
Overall, 
\begin{align*}
        &\sum_{j = 1}^{l} \sum_{i = 1}^{N-j}\bigg[\log p_j(d_j|\phib) + \log p_{j+i}(d_{j+i}|\phib)\bigg] =\\ 
        & \sum_{j = 1}^{l}(N - j)\log p_j(d_j|\phib) + \sum_{j = 1}^{l}(j-1) \log p_j(d_j|\phib) +\sum_{j = l+1}^{N}l \log p_j(d_j|\phib)=\\
        & \sum_{k=1}^{l}(N-1)\log p_k(d_k|\phib) +  \sum_{k=l+1}^{N}l \log p_k(d_k|\phib).
\end{align*}
Since $j \in \{1, \dots, l\}$ and
\begin{equation*}
b_{j + i} = 
   \begin{cases} 
      N - 1 & j + i\le l \\
      l & j + i > l 
   \end{cases},
\end{equation*}
the equality \ref{eqn:proposition:C} holds.

\paragraph{Proof of Proposition \ref{propostion:CvineTruncUnbiase}.}
The proof is identical with that of Proposition \ref{propostion:DvineTruncUnbiase} since each conditioned pair in the copula density $c_{j,(j+i); 1, \dots, (j-1)}$ is unique, and a C-vine is a special case of R-vine.

	\section{Simulation: Memory Profile}\label{app:mem}
	Here we present the memory profiles for the MH, the NUTS, and the Algorithm \ref{alg:TruncatedDfinal} under the simulation scenario studied in Chapter 5. These were recorded during a one hour period of running the algorithms. The MH algorithm and the NUTS were implemented in Python 3.0 using the PyMC3 module version 3.5. The memory profiles were measured using the memory-profiler module version 0.55.0 in Python 3.0. The VC was also implemented in Python 3.0. The code was run on the high performance computing cluster at the Institute for Cyber-Enabled Research at Michigan State University.
	
	\begin{figure}[H] 
		\begin{centering}
			\includegraphics[width=0.8\textwidth]{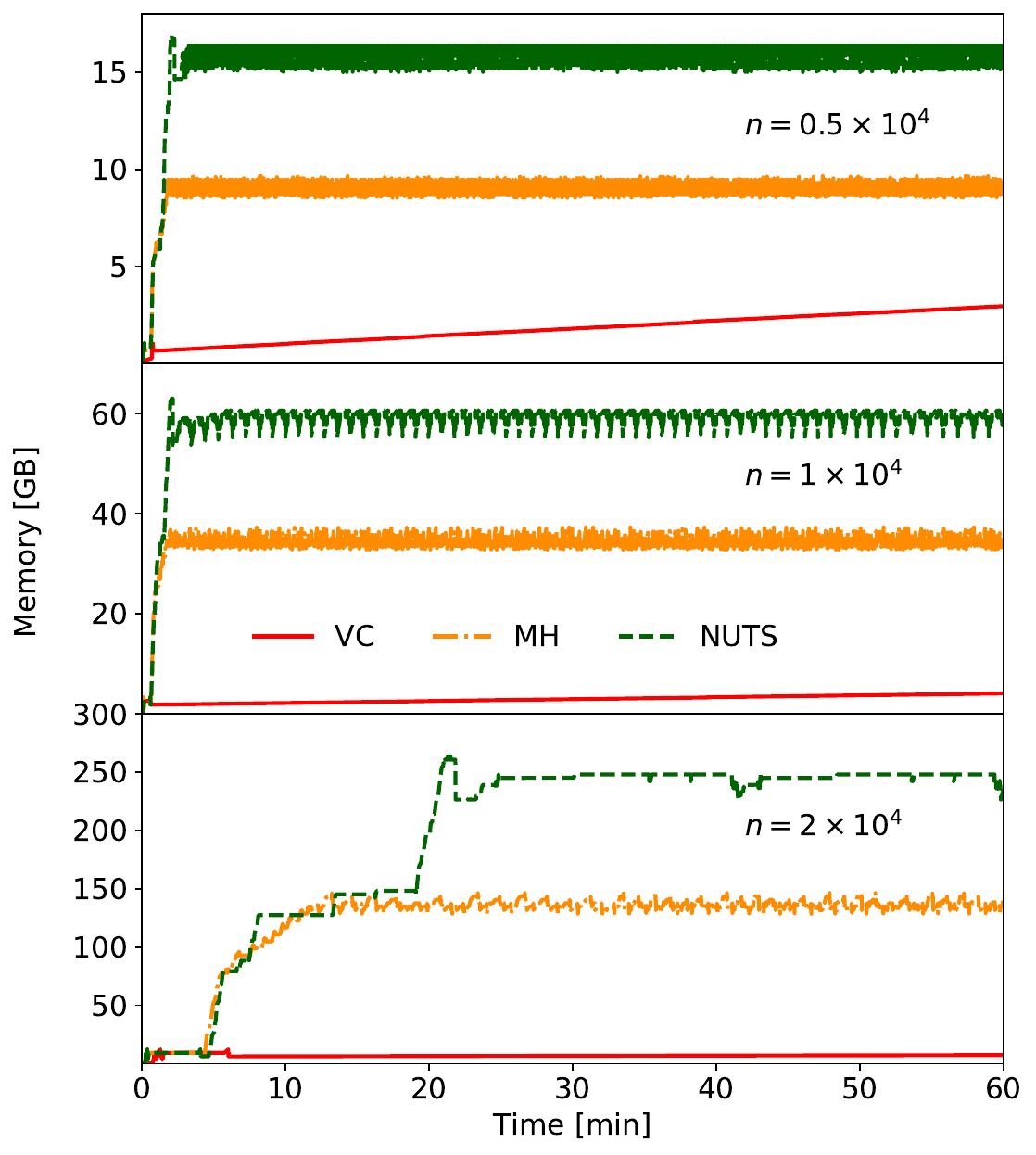}
			\caption{Recorded memory profiles of the Algorithm \ref{alg:TruncatedDfinal}, the MH algorithm, and the NUTS for the duration of 1h under the simulation scenario with $n = 0.5 \times 10 ^ 4$,  $n = 1 \times 10 ^ 4$, and  $n = 2 \times 10 ^ 4$.}
			\label{fig:mem}
		\end{centering}
	\end{figure}

	\section{Application: Liquid Drop Model (LDM)}\label{app:LDM}
	\subsection{GP specifications}
	In the case of the LDM $E_B(Z,N)$, we consider the GP prior with mean zero and covariance function
	
	\begin{align*}
	&\eta_f \times \text{exp}(-\frac{\|Z - Z' \|^2}{2\nu^2_Z} -\frac{\|N - N' \|^2}{2\nu^2_N} -\frac{\|\theta_{\rm vol} - \theta_{\rm vol}' \|^2}{2\nu^2_1} -\\
	& \frac{\|\theta_{\rm surf} - \theta_{\rm surf}' \|^2}{2\nu^2_2} -\frac{\|\theta_{\rm sym} - \theta_{\rm sym}' \|^2}{2\nu^2_3} -\frac{\|\theta_{\rm C} - \theta_{\rm C}' \|^2}{2\nu^2_4}).
	\end{align*}
	Similarly, we consider the GP process prior for the systematic discrepancy $\delta(Z,N)$ with mean zero and covariance function 
	\begin{equation*}
	\eta_\delta \times \exp{(-\frac{\|Z - Z' \|^2}{2l^2_Z} -\frac{\|N - N' \|^2}{2l^2_N})}.
	\end{equation*}
	\subsection{Experimental design}
	\cite{KoH} recommend to select the calibration inputs for the model runs so that any plausible value $\thetab$ of the true calibration parameter is covered. In this context, we consider the space of calibration parameters to be centered at the values of least squares estimates $\hat{\thetab}_{L_2}$ and broad enough to contain the majority of values provided by the nuclear physics literature \citep{Weizsacker1935,Bethe36,Myers1966,Kirson2008,Benzaid2020}. Table \ref{tab:LDMdesignParam} gives the lower and upper bounds for the parameter space so that $\text{Lower bound} = \hat{\theta}_{L_2} - 15 \times SE(\hat{\theta}_{L_2})$ and $\text{Upper bound} = \hat{\theta}_{L_2} + 15 \times SE(\hat{\theta}_{L_2})$. Here $SE(\hat{\theta}_{L_2})$ is given by the standard linear regression theory.
\setlength{\tabcolsep}{4pt}
\renewcommand{\arraystretch}{0.7}
\begin{table}[h]
\centering
		\begin{tabular}{l|ll}
			\hline 
			\hline \noalign{\smallskip}
			Parameter & Lower bound & Upper bound \\ \noalign{\smallskip}\hline\noalign{\smallskip}
			$\theta_{\rm vol}$ & 15.008 & 15.829 \\
			$\theta_{\rm surf}$ & 15.628 & 18.193 \\
			$\theta_{\rm sym}$ & 21.435 & 23.505 \\
			$\theta_{\rm C}$ & 0.665 & 0.72 \\ \noalign{\smallskip}\hline\hline 
		\end{tabular}
		\caption{The space of calibration parameters used for generating the outputs of semi-empirical mass formula \eqref{eqn:LDM}. \label{tab:LDMdesignParam}}
		\end{table}
	
	\subsection{Prior distributions}
	First, we consider the independent Gaussian distributions centered at the LS estimates $\hat{\thetab}_{L_2}$ (in Table \ref{tab:LDM_results})  with standard deviations $7.5 \times SE(\hat{\thetab}_{L_2})$ so that the calibration parameters used for generating the model runs are covered roughly within two standard deviations of the priors. Namely,
	\begin{align*}\label{eqn:priors_theta}
	\theta_{\textrm{vol}} &\sim \cN(15.42, 0.203), \\
	\theta_{\textrm{surf}} &\sim \cN(16.91, 0.645), \\
	\theta_{\textrm{sym}} &\sim \cN(22.47, 0.525), \\
	\theta_{\textrm{C}} &\sim \cN(0.69, 0.015).
\end{align*}
The prior distributions for hyperparameters of the GPs were selected as $\text{Gamma}(\alpha, \beta)$ with the shape parameter $\alpha$ and scale parameter $\beta$, so that they represent a vague knowledge about the scale of these parameters given by the literature on nuclear mass models \citep{Weizsacker1935,Bethe36,Myers1966, Fayans1998, Kirson2008, McDonnellPRL15, UNEDF0, UNEDF1, UNEDF2, Benzaid2020, kejzlar2020statistical}. In particular, the error scale $\sigma$ is in the majority of nuclear applications within units of MeV, therefore we set
\begin{equation*}
	\sigma \sim \text{Gamma}(2,1),
\end{equation*}
with the scale of the systematic error being
\begin{equation*}
	\eta_\delta \sim \text{Gamma}(10,1),
\end{equation*}
to allow for this quantity to range between the units and tens of MeV. It is also reasonable to assume that the mass of a given nucleus is correlated mostly with its neighbours on the nuclear chart. We express this notion through these reasonably wide prior distributions

	\begin{align*}
	l_Z &\sim \text{Gamma}(10,1), \\
	l_N &\sim \text{Gamma}(10,1), \\
	\nu_Z &\sim \text{Gamma}(10,1), \\
	\nu_N &\sim \text{Gamma}(10,1), \\
	\nu_i &\sim \text{Gamma}(10,1), \hspace{1cm} i = 1,2,3,4.
	\end{align*}

Finally, the majority of the masses in the training dataset of 2000 experimental binding energies fall into the range of $[1000, 2000]$ MeV (1165 of masses precisely). We consider the following prior distribution for the parameter $\eta_f$ to reflect on the scale of the experimental binding energies:
\begin{equation*}
	\eta_f \sim \text{Gamma}(110,10).
\end{equation*}

\end{appendices}

\bibliographystyle{spbasic}      
\bibliography{biblio} 
\end{document}